\documentclass[
    aps,
    pra,
    amsmath,
    amssymb,
    11pt,
    onecolumn,
    nofootinbib,
    superscriptaddress
]{revtex4-2}

\usepackage[T2A]{fontenc}
\usepackage[utf8x]{inputenc}
\usepackage[russian,english]{babel}

\usepackage{graphicx}
\usepackage{longtable}
\usepackage{array}

\usepackage[
    a4paper,
    left=30mm,
    right=30mm,
    top=25mm,
    bottom=25mm
]{geometry}

\input{maik_edt.rty}

\def\be{\begin{equation}}
\def\ee{\end{equation}}
\def\ba{\begin{eqnarray}}
\def\ea{\end{eqnarray}}

\def\12{{1\over 2}}

\def\ltsima{$\; \buildrel < \over \sim \;$}
\def\simlt{\lower.5ex\hbox{\ltsima}}
\def\gtsima{$\; \buildrel > \over \sim \;$}
\def\simgt{\lower.5ex\hbox{\gtsima}}

\newcommand{\hh}{$\rm H_2$}
\newcommand{\CI}{C$\,${\sc i}}

\newcommand{\SiII}{Si$\,${\sc ii}}
\newcommand{\FeII}{Fe$\,${\sc ii}}
\newcommand{\Lya}{Ly$\alpha$}
\newcommand{\Ha}{H$\alpha$}

\newcommand\ion[2]{\text{#1\,\textsc{\lowercase{#2}}}}
\newcommand{\ionpwne}[2]{#1\,{\sc #2}}

\newcommand{\lineWL}[2]{#1~$\lambda$#2~\AA}

\newcommand{\HeIIpwne}{\ionpwne{He}{ii}}
  
\newcommand{\CIIpwne}{\ionpwne{C}{ii}} \newcommand{\CIIp}{[\CIIpwne]}
\newcommand{\CIIIpwne}{\ionpwne{C}{iii}} \newcommand{\CIIIp}{[\CIIIpwne]}
\newcommand{\CIVpwne}{\ionpwne{C}{iv}} \newcommand{\CIVp}{[\CIVpwne]}

\newcommand{\NIIIpwne}{\ionpwne{N}{iii}} \newcommand{\NIIIp}{[\NIIIpwne]}
\newcommand{\NIVpwne}{\ionpwne{N}{iv}} \newcommand{\NIVp}{[\NIVpwne]}
 
\newcommand{\OIIpwne}{\ionpwne{O}{ii}} \newcommand{\OIIp}{[\OIIpwne]}
\newcommand{\OIIIpwne}{\ionpwne{O}{iii}} \newcommand{\OIIIp}{[\OIIIpwne]}

\newcommand{\NeIVpwne}{\ionpwne{Ne}{iv}} \newcommand{\NeIVp}{[\NeIVpwne]}
\newcommand{\HaLine}{$\text{H}_\alpha$}

\newcommand{\La}{$\text{Ly}_\alpha$}
\newcommand{\LaWL}{\lineWL{\La}{1215}}
\newcommand{\HeIIlineShort}{\lineWL{\HeIIpwne}{1215}}
\newcommand{\HeIIlineLong}{\lineWL{\HeIIpwne}{1640}}
\newcommand{\CIIlineShort}{\lineWL{\CIIpwne}{1335}}
\newcommand{\CIIlineLong}{\lineWL{\CIIp}{2326}}
\newcommand{\CIIIline}{\lineWL{\CIIIp}{1908}}
\newcommand{\CIVline}{\lineWL{\CIVpwne}{1550}}
\newcommand{\NIIIline}{\lineWL{\NIIIp}{1750}}
\newcommand{\NIVline}{\lineWL{\NIVp}{1487}}
\newcommand{\OIIlineShort}{\lineWL{\OIIp}{2470}}
\newcommand{\OIIlineLong}{\lineWL{\OIIp}{3727}}
\newcommand{\OIIIline}{\lineWL{\OIIIp}{1663}}
\newcommand{\NeIVline}{\lineWL{\NeIVp}{2424}}

\begin{document}
\selectlanguage{english}

\noindent {\it ASTRONOMY REPORTS, 2026, accepted}
\bigskip\bigskip  \hrule\smallskip\hrule
\vspace{15mm}

\title{\bf Interstellar and intergalactic medium in the ultraviolet band}
\author{\bf\firstname{M.~S.}~\surname{Kirsanova}}
\email{kirsanova@inasan.ru}
\affiliation{Institute of Astronomy of the Russian Academy of Sciences, Moscow, Russia}

\author{\bf\firstname{S.~A.}~\surname{Balashev}}
\affiliation{Ioffe Institute of Physics and Technology, St. Petersburg, Russia
}

\author{\bf\firstname{M.~V.}~\surname{Barkov}}
\affiliation{Institute of Astronomy of the Russian Academy of Sciences, Moscow, Russia}

\author{\bf\firstname{Е.~О.}~\surname{Vasiliev}}
\affiliation{P.N. Lebedev Physical Institute of the Russian Academy of Sciences, Moscow, Russia}

\author{\bf\firstname{D.~S.}~\surname{Wiebe}}
\email{dwiebe@inasan.ru}
\affiliation{Institute of Astronomy of the Russian Academy of Sciences, Moscow, Russia}
\affiliation{Department of Chemistry, Lomonosov Moscow State University, Moscow, Russia}

\author{\bf\firstname{I.~I.}~\surname{Zinchenko}}
\affiliation{A. V. Gaponov-Grekhov Institute of Applied Physics of the Russian Academy of Sciences, Nizhny Novgorod, Russia}

\author{\bf\firstname{А.~V.}~\surname{Moiseev}}
\affiliation{Special Astrophysical Observatory of the Russian Academy of Sciences, Nizhny Arkhyz, Russia}

\author{\bf\firstname{M.~S.}~\surname{Murga}}
\affiliation{Institute of Astronomy of the Russian Academy of Sciences, Moscow, Russia}

\author{\bf\firstname{I.~N.}~\surname{Nikonorov}}
\affiliation{Institute of Astronomy of the Russian Academy of Sciences, Moscow, Russia}

\author{\bf\firstname{M.~E.}~\surname{Sachkov}}
\affiliation{Institute of Astronomy of the Russian Academy of Sciences, Moscow, Russia}

\author{\bf\firstname{I.~S.}~\surname{Khrykin}}
\affiliation{Pontifical Catholic University of Valparaíso, Valparaíso, Chile}

\author{\bf\firstname{B.~M.}~\surname{Shustov}}
\affiliation{Institute of Astronomy of the Russian Academy of Sciences, Moscow, Russia}

\author{\bf\firstname{Yu.~A.}~\surname{Schekinov}}
\affiliation{Raman Research Institute, Bangalore, India}

\author{\bf\firstname{A.~D.}~\surname{Yarovova}}
\affiliation{Sternberg Astronomical Institute, Lomonosov Moscow State University, Moscow, Russia}

\begin{abstract}
\vspace{3mm}
\vspace{3mm}

This article presents an overview of key scientific challenges in the astrophysics and astrochemistry of the interstellar and intergalactic media. Addressing many of these challenges will become possible with the Russian space observatory Spektr-UF, which is currently being prepared for launch. The ultraviolet (UV) spectral range is critically important for studying the evolution of the Universe, as it contains fundamental absorption and emission lines of most chemical elements and the most important molecules. These lines make it possible to reliably determine the physical conditions and chemical composition of various astrophysical objects, as well as the characteristics of the processes occurring within them. The article examines the prospects for UV observations across a broad range of scientific problems, from investigating the epoch of helium reionization and the thermal history of the Universe to mapping ongoing star formation in galaxies. Dedicated sections address the study of the chemical evolution of galaxies at intermediate redshifts, the distribution of “hidden” baryons and metals, the small-scale structure of interstellar clouds, the physics of photodissociation regions and protoplanetary disks, and the modeling of UV emission from nebulae surrounding fast-moving pulsars. The high sensitivity and spectral resolution of the Spektr-UF instruments are expected to provide fundamentally new data across all these research areas, filling existing gaps in our understanding of the interstellar and intergalactic media.

\end{abstract}

\maketitle

\section{Introduction}\label{vved} 

Ultraviolet radiation is one of the key factors governing the evolution of the Universe. It largely determines the physical properties of the diffuse medium in the Universe, its molecular composition, the star formation rate, and the properties of newly formed stars. In the present-day Universe, the main sources of UV radiation in galaxies are massive O- and B-type stars. Their photospheric temperatures exceed 10,000 K, and over their short lifetimes of several million to tens of millions of years, these stars produce enormous fluxes of ionizing photons. White dwarfs, cataclysmic variables, and shock waves generated by supernova explosions also contribute to the overall UV radiation field. Such a field is present even in low-density regions of the diffuse interstellar medium, where the background UV radiation is sustained by the scattering of starlight by interstellar dust.

In the extragalactic medium, the main sources of UV photons at high redshifts were young dwarf galaxies undergoing intense star formation and, even more importantly, accreting supermassive black holes—quasars. Quasars are extremely luminous in the UV and can ionize hydrogen on megaparsec scales. Today, the main source of UV photons in the voids between galaxies is the cosmic ultraviolet background—the combined radiation from quasars and galaxies.

The interstellar medium (ISM) is an important, complex, and highly diverse component of galaxies that remains insufficiently understood because of limited observational capabilities. A wide variety of processes take place in the ISM, ranging from microscopic scales, including the formation of cosmic dust and molecules and the heating and cooling of gas, to global scales involving the formation of stars and planets. On still larger scales, the ISM plays a key role in shaping the formation and evolution of galaxies as a whole (e.g., \cite{Maiolino2019,Peroux2020}).

The interstellar medium is affected by numerous processes that regulate its physical conditions, spatial structure, dust content and properties, metallicity, and relative elemental abundances, ultimately giving rise to a complex and dynamic system.

Gas in the ISM has a complex multiphase structure, with its density and temperature spanning extremely broad ranges. The structure of the gas in the ISM is intrinsically linked to the so-called baryon cycle: at the final stages of their evolution, stars inject substantial amounts of energy and newly synthesized heavy elements into the ISM through stellar winds and supernova explosions. In addition to strongly ionizing the medium and producing a hot ionized phase, these processes drive an active exchange of heavy elements between the gas phase and the surfaces of dust grains. Dust grains form in stellar winds and supernova ejecta, but are also destroyed in the hot gas behind shock fronts.

This hot gas fills a substantial fraction of the ISM volume and, as it gradually cools, partly through radiative cooling in heavy-element lines, gives rise to the warm ionized medium and, following recombination, to the neutral medium.

Further cooling of the neutral phase through emission in the fine-structure lines of heavy elements, such as OI, CII, and SiII, leads to the formation of the cold phase of the ISM. In this phase, the formation of molecular hydrogen on the surfaces of dust grains, as well as that of other molecules, including CO, becomes efficient. These molecules promote further rapid cooling of the medium to temperatures of only a few kelvin. The subsequent condensation of the cold ISM to densities several orders of magnitude higher than the mean density of the diffuse medium leads to the formation of protostellar condensations and protoplanetary disks, where the process of forming new stars begins.

Thus, the ISM is a complex and continuously evolving system that not only governs the process of star formation but also preserves a record of its history in its chemical and physical state. Current observations show that the ISM is highly inhomogeneous in both its physical properties and chemical composition, exhibiting a clumpy and complex structure that evolves with time (see, e.g., \cite{Draine2011}). Observations with the Spektr-UV telescope can substantially improve our understanding of the composition and physical properties of the ISM, as well as its inhomogeneities.

\begin{figure*}
    \includegraphics[width=0.95\linewidth]{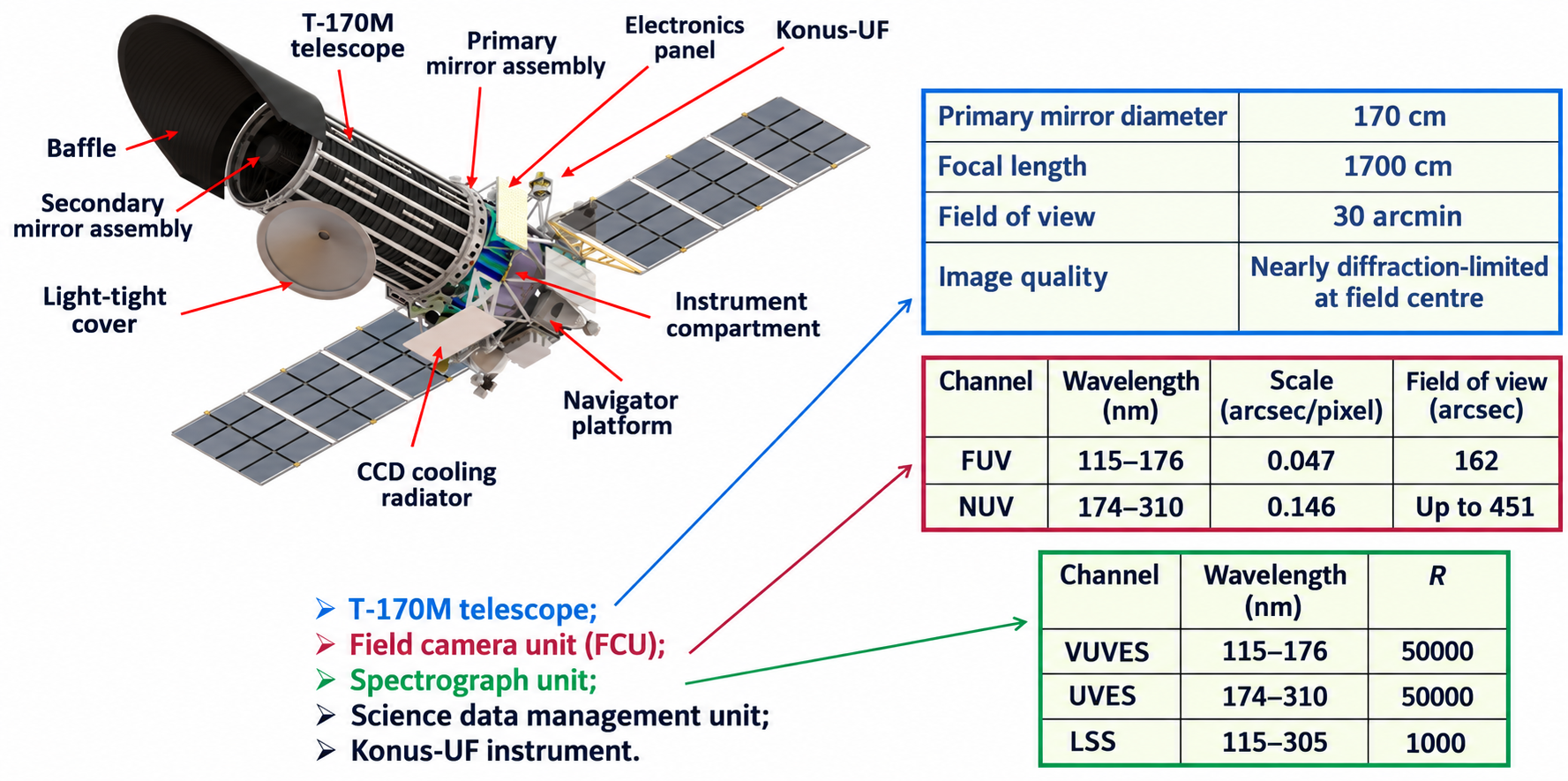}
    \caption{Russian space observatory Spektr-UF}
    \label{fig:Spektr_UF}
\end{figure*}

The Russian space observatory Spektr-UV is currently under development~\cite{Shustov_2018, Shustov_2021} and is designed to study a wide variety of astrophysical objects and phenomena in the Universe~\cite{Boyarchuk_2016}. The observatory will be equipped with far- and near-UV spectrographs~\cite{Shugarov_2023}, as well as a field camera unit~\cite{Shmagin_2023}. A schematic view of the observatory is shown in Fig.~\ref{fig:Spektr_UF}. Preliminary estimates of the observing time required with these instruments can be obtained using the exposure time calculator~\cite{2024INASR...9...12T}.

This paper reviews the current state of UV studies of the ISM and the intergalactic medium (IGM) and discusses the scientific questions that can be addressed with the instruments aboard Spektr-UF.

\section{Intergalactic and interstellar medium in galaxies}

\subsection{UV Observations of Molecular Hydrogen in the Universe}

Molecular hydrogen \hh\ is the most abundant molecule in the Universe (its abundance exceeds that of other molecules by $\gtrsim10^4$). The emission from this molecule can only be observed under rather specific conditions. This is due to two factors. First, \hh\ is a homonuclear molecule and therefore has no dipole moment in its electronic ground state, which leads to the suppression of rovibrational transitions. Second, the rovibrational transitions of \hh\ lie in the infrared range, as opposed to the more observationally convenient radio range. The emission from the \hh\ molecule can be efficiently observed in the UV range using transitions between the ground electronic state ($\rm X^1\Sigma^+_g$) and the first two excited electronic states ($\rm B^1\Sigma^+_u$ and $\rm С^1\Pi_u$). The sets of lines corresponding to these \hh\ transitions are known as the Lyman bands ($\rm B^1\Sigma^+_u \to X^1\Sigma^+_g$) and Werner bands ($\rm C^1\Pi_u \to X^1\Sigma^+_g$). These transitions have characteristic wavelengths (for the lowest vibrational level) $\lambda_{\rm H2}\lesssim1150$\,\AA, i.e., they lie in the far-UV range. Since H$_2$ molecules are found predominantly in cold molecular clouds, these transitions can be observed in absorption along lines of sight to bright background sources (although fluorescent emission has also been observed in photodissociation regions and protoplanetary disks).

\begin{figure*}
	\includegraphics[width=0.95\linewidth]{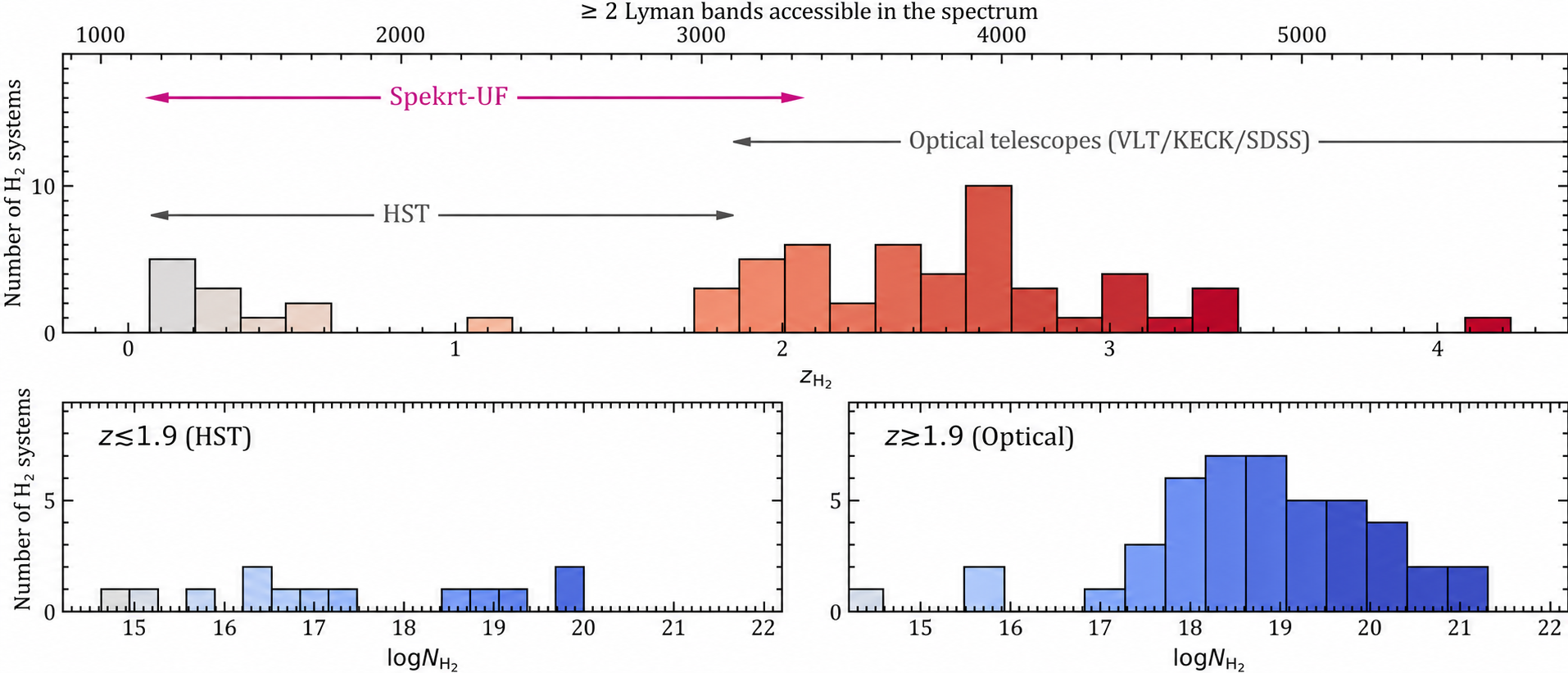} 
	\caption{Statistics of known \hh\ absorption systems in the Universe. The upper panel shows their distribution as a function of redshift, with lines indicating the wavelength ranges accessible to optical and space-based telescopes, including Spektr-UF. The upper horizontal axis of this panel indicates the wavelength above which at least the first two Lyman bands become accessible, which are conventionally required for a detailed analysis of \hh\ lines. The lower right and left panels show the distributions of line-of-sight column densities for systems identified using optical and space-based telescopes, respectively.}
	\label{fig:h2_stats}
\end{figure*}

The Lyman and Werner bands of \hh\ were first detected in the molecular clouds of our Galaxy along lines of sight to bright nearby stars using rocket-borne observations \cite{Carruthers1970}, and were subsequently studied in detail by the Copernicus UV Observatory \cite{Spitzer1974} and in even greater detail by the FUSE (Far Ultraviolet Spectroscopic Explorer) observatory \cite{Sahnow1996}. In total, approximately 140 absorption systems associated with our Galaxy \cite{Shull2021} and about 100 in the Magellanic Clouds \cite{Welty2012} have been identified. The wavelength coverage of the Spektr-UF telescope starts at 1150\AA, making direct studies of molecular hydrogen in the {\em local} Universe impossible, except for specific investigations involving observations of highly excited \hh\ levels. Nevertheless, the telescope's high sensitivity and spectral resolution will enable studies of the interstellar medium through observations of \hh\ at intermediate redshifts of $z= 2-3$. Indeed, due to cosmological redshift, molecular hydrogen lines are shifted toward longer wavelengths. At $z\gtrsim2$, \hh\ lines fall in the optical range, allowing \hh\ absorption systems to be studied at high spectral resolution using the world's largest optical telescopes, such as the VLT (Very Large Telescope) and Keck\footnote{Although the first identification of molecular hydrogen \cite{Levshakov1985} was made before the VLT and Keck became operational, most systems observed in the optical range were discovered using these telescopes (see, e.g., \cite{Balashev2019} and references therein)}. In these observations, the background sources are cosmologically distant quasars and gamma-ray burst afterglows. Targeted programs investigating damped Lyman-$\alpha$ (DLA) systems and searching for \hh\ \cite{Ledoux2003,Noterdaeme2008,Jorgenson2014} with the largest optical telescopes have yielded a sample of $\sim 60$ \hh\ absorption systems in the distant Universe. An important milestone was the development of methods for selecting \hh\ absorption systems using the SDSS spectroscopic survey \cite{Srianand2008,Balashev2014,Noterdaeme2018,Noterdaeme2019}. Considerably fewer molecular hydrogen absorption systems have been identified at intermediate redshifts of $z\lesssim2$, primarily through studies using archival quasar observations obtained with the Hubble Space Telescope (hereafter HST) \cite{Muzahid2016}. Fig. \ref{fig:h2_stats} shows the distribution of identified distant \hh\ absorption systems in quasar spectra as a function of redshift and line-of-sight column density.

It can be seen that the intermediate-redshift range remains poorly explored. Most systems identified by HST at redshifts of $z\lesssim2$ have been studied using the Cosmic Origins Spectrograph (COS) spectrograph, which has a maximum spectral resolution of $\approx20000$. Spektr-UF will have a higher spectral resolution, comparable to that of optical telescopes, enabling more detailed studies of \hh\ in molecular clouds. In addition, only a few \hh\ absorption systems with high column densities ($\log N (\rm H_2) \gtrsim19$, where $N$ is measured in cm$^{-2}$) are known at $z\lesssim 2$. Such systems are of particular interest because they trace star-forming regions and allow other important species in the cold phase of the interstellar medium (CI, CII, and the molecules HD and CO) to be identified\footnote{The lines of these atoms and molecules lie in the same or a more convenient, longer-wavelength region of the spectrum than \hh\ and therefore generally do not require additional observations.}, as well as enabling detailed assessments of the physical conditions in this phase.

Before describing in greater detail several observational projects involving \hh\ that can be pursued with Spektr-UF, we note that high-resolution spectra ($R\sim 50000$) are required for a reliable analysis of the lines. Somewhat paradoxically, absorption systems at high redshifts have been studied in greater detail than those at intermediate redshifts and in the local Universe, since only one telescope (HST) can perform high-resolution UV spectroscopy, whereas numerous powerful ground-based optical telescopes are available. The spectrographs aboard Spektr-UF can therefore make a significant contribution to \hh\ observations.

\subsection{Physical Conditions in the Cold Phase of the Interstellar Medium}
\label{sec:physc_cond_cold_ISM}

Under typical interstellar medium conditions, \hh\ is predominantly found in the cold phase of the ISM, namely, in molecular clouds \cite{Balashev2019}. This is because relatively high gas densities, $n \gtrsim 10^2$\,cm$^{-3}$, are required for molecular hydrogen to <<survive>> in a typical UV radiation field. Since stars are thought to form in the cold, dense phase of the ISM, studying its physical conditions and chemical composition is essential for understanding the star formation process. Observations of \hh\ lines and associated species provide a unique opportunity to investigate these properties. In particular, \hh\ is usually detected through transitions involving several rotational levels of its ground vibrational state. These levels are populated through collisions and radiative pumping \cite{Abgrall1992} via excited electronic states; therefore, their populations depend on the gas density, temperature, and intensity of the UV radiation field. This makes it possible to estimate these parameters by modelling the observed rotational-level populations \cite{Klimenko2020}. Because radiative pumping involves resonant transitions, its modelling requires detailed calculations of radiative transfer in \hh\ lines, as implemented in codes such as {\sc Cloudy} \cite{Ferland2017} and {\sc PDR Meudon} \cite{LePetit2006}. However, applications of these codes to both individual systems \cite{Noterdaeme2017,Balashev2019} and systematic samples \cite{Klimenko2020} have shown that incorporating the populations of the \CI\ fine-structure levels is also desirable to break parameter degeneracies \cite{Kosenko2024}. Furthermore, the relative abundance of \CI/\hh\ can be used to constrain an important parameter of the ISM: the cosmic-ray ionization rate \cite{Balashev2024}.
 
Another way to constrain the physical conditions is to identify the molecules HD and CO associated with \hh. While observations of HD (in combination with \hh\ and \CI) can be used to constrain the cosmic-ray ionization rate \cite{Balashev2020,Kosenko2021,Kosenko2023}, the abundance of CO and the populations of its rotational levels provide information about the gas density in the most strongly shielded part of an absorption system\footnote{Note that this is generally true for the local environment and for systems at intermediate redshifts of $z\lesssim1.5$, since at higher redshifts the CO rotational levels in diffuse gas are populated predominantly by the cosmic microwave background radiation \cite{Noterdaeme2011}.}. In addition, ions such as Cl$^+$ and Ar$^+$ actively participate in chemical reactions with \hh\ in the interstellar medium (see, e.g., \cite{Noterdaeme2023}); therefore, measuring their abundances makes it possible to obtain independent estimates of the molecular fraction of the gas \cite{Balashev2015}.

Recent systematic measurements of the physical conditions in \hh\ absorption systems reveal a systematic evolution of the thermal state of the cold ISM with redshift, despite substantial scatter in the measurements \cite{Balashev2023}. This trend may reflect the influence of metallicity on the thermal balance of the gas \cite{Bialy2019,Balashev2022} and the evolution of metallicity with redshift. Meanwhile, the cosmic star formation rate density declined by approximately an order of magnitude between $z=2$ and $z=0$, suggesting a corresponding decrease in the {\em local} UV radiation field in the ISM. Observations of \hh\ and associated species provide a means of constraining this radiation field. Together, these trends motivate high-resolution observations of \hh\ absorption systems at intermediate redshifts, which can shed light on the evolution of the physical conditions in the cold ISM.  

\subsection{Molecular Gas in the Vicinity of Active Galactic Nuclei}

A recent systematic analysis of SDSS spectra revealed a population of \hh\ absorption systems located close to quasars in redshift space \cite{Noterdaeme2019}. The incidence of \hh\ detection in these systems is at least several times higher than in systems separated from the quasars in redshift, contrary to the simplest expectation that the intense UV radiation field of a quasar should promote the photodissociation of \hh. This excess of \hh\ absorbers may instead reflect the clustering of gas and galaxies around quasar host galaxies. Supporting this interpretation, \Lya\ emission has been detected at the centre of saturated absorption lines in approximately 50 per cent of \hh\ systems close to quasars. Subsequent observations \cite{Noterdaeme2023} showed that this emission originates from regions extending over $\gtrsim10$\,kpc around the quasar.

In addition, \hh\ absorption systems may be associated with galactic outflows \cite{Noterdaeme2021}, which frequently contain molecular gas \cite{Cicone2014,Veilleux2013}, or with companion galaxies undergoing mergers with quasar host galaxies \cite{Balashev2025}. Importantly, modelling the observed \hh\ level populations provides a diagnostic of the surrounding environment, enabling constraints on the gas density and temperature, as well as estimates of the distance between the absorber and the quasar (see, e.g., \cite{Noterdaeme2021,Balashev2025}). Such diagnostics are difficult to obtain using most alternative methods, particularly at high redshifts, where outflows are more common and more extended, owing to the greater energy output of active galactic nuclei. In these systems, the gas is generally spatially unresolved with existing telescopes. Observations of \hh\ absorption systems at intermediate redshifts with the Spektr-UF spectrograph would enable detailed studies of molecular gas around AGN in the relatively nearby Universe. The lower redshifts ($z\lesssim1$) of these systems also make fainter objects accessible and provide substantially better spatial resolution for follow-up observations of their emission lines. This would enable studies of both the host galaxies and the physical conditions and mass of the gas surrounding AGN (see, e.g., \cite{VillaVelez2024}).

The ability to study absorption systems at intermediate redshifts also enables detailed investigations of the galaxies associated with \hh\ absorbers. Recent studies have begun to explore the fields surrounding quasars with detected \hh\ absorption systems \cite{Zahedy2020,Boettcher2021}, offering a way to establish the origin of the absorbing gas. Although the inferred physical conditions indicate that this gas resides in the diffuse interstellar medium \cite{Balashev2023}, the identity of the host galaxies remains uncertain. Because absorption-line surveys toward background sources select systems according to their absorption cross-section, with larger cross-sections increasing the probability of detection, statistical studies of the associated galaxies could establish how diffuse molecular gas is distributed among different galactic environments.

Absorption-line studies can also probe structures surrounding galaxies, including gas in the tidal features of interacting systems. Current observations of galaxies in the fields surrounding absorption systems are conducted primarily in the optical, using imaging or integral-field spectrographs such as MUSE and KCWI, and at millimetre wavelengths, using facilities such as NOEMA and ALMA. These observations provide information on stellar populations and ionized gas, and on dense molecular gas, respectively. The advent of next-generation radio interferometers, including MeerKAT and ASKAP, is opening up opportunities to study atomic gas in these galaxies \cite{Maina2022}, while the SKA will extend such studies across the full intermediate-redshift range. This science case is not limited to \hh-bearing absorbers: it also applies to classical DLA systems, for which the key transitions of heavy elements at $z\lesssim1$ will largely fall within the wavelength coverage of Spektr-UF.

\subsection{Gamma-ray burst afterglows}

The absorption studies described above use quasars as background sources. Their persistence, despite intrinsic variability, and high brightness make them well suited to obtaining high-quality spectra for absorption-line analysis. An alternative class of background sources is provided by gamma-ray burst (GRB) afterglows: emission following the initial gamma-ray flash that fades over several hours at longer wavelengths, particularly in UV. A key feature of GRB afterglow spectra is the almost ubiquitous presence of absorption lines from neutral and ionized gas in the interstellar and circumgalactic media of the host galaxy. These transitions, most of which lie in the UV, provide diagnostics of the chemical composition and physical conditions of the absorbing gas \cite{Perley2016}.

Because GRBs occur in the central regions of distant, actively star-forming galaxies, the detection rate of \hh\ absorption reaches 40 per cent, substantially exceeding that in DLA systems identified in quasar spectra ($\approx 4$ per cent; \cite{Balashev2018}). Some afterglows also reveal the effects of GRB radiation on the host-galaxy gas, including variable absorption lines \cite{DeCia2012} and enhanced populations of fine-structure levels in ions such as \SiII\ and \FeII. Molecular hydrogen is therefore an especially valuable probe of the physical conditions in interstellar gas exposed to an intense, time-dependent radiation field \cite{Draine2002}. Absorption-line observations are, however, limited by the unpredictable occurrence of GRBs, the rapid fading of their afterglows and the lack of opportunities for repeat observations. Statistical samples are thus essential for a comprehensive understanding of their host galaxies. Current and planned programmes provide near-instantaneous burst localizations and basic properties, enabling systematic studies of GRB hosts.

Statistical studies of GRB afterglows show that the observed GRB redshift distribution peaks at $z\sim 1$. Optical telescopes such as VLT, however, can probe host-galaxy absorption spectra in detail only at $z\gtrsim2$, where fainter afterglows and shorter observing windows substantially limit spectral quality. The wavelength coverage of Spektr-UV will make absorption-line studies possible for most GRB afterglows and enable high-resolution spectroscopy. By comparison, optical studies have predominantly relied on the medium-resolution VLT/X-shooter spectrograph.

The unpredictable occurrence and sky locations of GRBs require rapid telescope slewing and dedicated transient-observation programmes that enable observations before the afterglow fades.

\subsection{Helium reionization and quasar properties}

The epoch of reionization ($z \lesssim 6$), during which the IGM became fully ionized, marks the last major cosmological phase transition. It is conventionally divided into two distinct periods: hydrogen reionization, thought to have ended by $z\simeq 5-6$ \cite{Fan2006a, Dayal2018, Gnedin2022}, and helium reionization, which was complete by $z\simeq 2.7$ \cite{McQuinn2009, Worseck2016, LaPlante2017}. Establishing the timing and spatial structure of reionization is essential to understanding cosmic evolution and constraining the properties and evolution of the first stars, galaxies and large-scale structure \cite{mcquinn2016}.

In the prevailing picture, the complete reionization of intergalactic helium---the removal of both electrons---occurred in two distinct stages
. During the first stage, ultraviolet radiation from stellar populations in galaxies singly ionized most neutral helium atoms ($\ion{He}{I} \rightarrow \ion{He}{II}$, $h\nu \gtrsim 24.6~{\rm eV}$), nearly coincident with the ionization of intergalactic hydrogen ($\ion{H}{I} \rightarrow \ion{H}{II}$, $h\nu \gtrsim 13.6~{\rm eV}$) by $z \sim 6$ \cite{FG2008, Eide2020}. Stars, however, produce too few photons with energies $h\nu \gtrsim 54.4~{\rm eV}$ to fully ionize helium atoms ($\ion{He}{II} \rightarrow \ion{He}{III}$; hereafter, \ion{He}{II} reionization). Quasars are the leading candidates for supplying this hard ionizing radiation to the IGM, as supported by measurements of their spectral energy distributions and luminosity functions \cite{Hopkins2007,Willott2010}. Intergalactic helium therefore remained predominantly singly ionized until $ 2.5 \lesssim z \lesssim 4.0$, when the quasar population became sufficiently abundant to supply the required photons with energies $h\nu \gtrsim 54.4~{\rm eV}$ \cite{LaPlante2017, Furlanetto2008, Kulkarni2019}.

The only direct probe of \ion{He}{II} reionization is \ion{He}{II} Lyman-$\alpha$ absorption ($\lambda_{\rm rest} = 303.78$\AA) in quasar spectra ($z_{\rm em} \gtrsim 2$), which requires sufficiently sensitive space-based telescopes. Spectra obtained with FUSE and HST/COS reveal a highly inhomogeneous reionization process \cite{Worseck2016, Worseck2011, Syphers2012, Worseck2019}. Much as \ion{H}{I} Lyman-$\alpha$ absorption probes the end of hydrogen reionization at $z \gtrsim 5-6$, the \ion{He}{II} Lyman-$\alpha$ forest observed towards quasars at $z_{\rm em} \sim 2.7$ traces the final stages of \ion{He}{II} reionization. At higher redshifts, the forest gives way to regions of complete absorption, indicating a greater abundance of singly ionized helium.

A statistical analysis of the \ion{He}{II} effective optical depth and transmission spikes in high-resolution HST/COS spectra ($R = \lambda / \Delta \lambda \sim 12500-18000$) spanning $ 2.3 \lesssim z \lesssim 3.8$ \cite{Worseck2016} revealed a gradual increase in effective optical depth from $\tau \simeq 1$ to $\tau \gtrsim 4.5$, together with substantial sightline-to-sightline scatter. This behaviour is consistent with models in which intergalactic helium is predominantly fully ionized and the ionizing background exhibits large spatial fluctuations \cite{Worseck2019, Makan2021}. Photon energy in excess of the $54.4~{\rm eV}$ ionization threshold heats the IGM, substantially affecting the properties of the \ion{H}{I} Lyman-$\alpha$ forest at $z\simeq 2-3$ \cite{Khrykin2017}. Constraining the ionizing background and IGM temperature during and after helium reionization is therefore essential for hydrodynamical models of the evolution of large-scale structure \cite{Davies2017, Onorbe2017, Puchwein2019}.

The declining abundance of quasars at high redshift ($z \gtrsim 3$), together with absorption by intervening \ion{H}{I} Lyman-limit systems ($\lambda \lesssim 912$\AA) at lower redshifts, severely restricts the sample of sufficiently bright quasars suitable for \ion{He}{II} Lyman-$\alpha$ absorption studies to a few dozen objects. Expanding this sample requires an initial search for quasars at $z_{\rm em} \gtrsim 3$. Candidates can be drawn from archival surveys such as SDSS ($\sim 40000$ identified quasars) and newer surveys such as DESI ($\sim 90000$ identified quasars at the time of writing), then cross-matched with the GALEX ultraviolet survey. The sensitivity of the Spektr-UV/UVES spectrograph \cite{Shugarov_2023} will enable observations of quasars at $z_{\rm em} \simeq 4$, opening a window onto the early stages of helium reionization.

\subsection{Indicators of ongoing star formation in galaxies} \label{sec:moiseev}

Processes associated with star formation, including supernova explosions, winds from massive stars and radiation pressure, are key drivers of the baryon cycle in the interstellar medium of galaxies. Their role is particularly important in low- and intermediate-mass galaxies, where active galactic nuclei have a comparatively modest influence on the ISM \cite{Donahue2022}. Quantifying the energy supplied to the ISM by young stellar populations across different galaxy types requires measurements of the current star formation rate (SFR). Conventional SFR estimates use luminosities in spectral bands that trace emission from young stars (blue optical and ultraviolet photometry), the response of the ISM to ionizing photons from OB stars (optical emission lines such as \Ha{} or [OII]$\lambda3727$), dust heated by young stars (mid- and far-infrared emission), or the remnants of recent supernova explosions (X-ray and non-thermal radio emission). Reviews of these indicators and their calibrations are provided by \cite{Kennicutt1998ARA&A..36..189K,Kennicutt2012ARA&A..50..531K}. Different SFR indicators probe timescales set by the main-sequence lifetimes of stars of different masses. Comparing them therefore provides insight into the star formation history, as demonstrated in the pioneering study of irregular galaxies by \cite{Gallagher1984ApJ...284..544G}.

Over the past two decades, \Ha{} emission and space-based ultraviolet measurements have become two of the most widely used tracers of ongoing star formation; the corresponding estimates are denoted here as SFR(\Ha) and SFR(UV). Their popularity reflects both the availability of observations---including the extensive GALEX ultraviolet survey \cite{2007ApJS..173..682M}, narrow-band \Ha{} imaging and integral-field spectroscopy---and the ability to map star formation at relatively high angular resolution: $4-6''$ for SFR(UV) and $1-2''$ for SFR(\Ha). Reconstructing the star formation history from these tracers relies on their different temporal sensitivities. SFR(\Ha) is governed by stars with masses $m>10\,M_\odot$ and traces a characteristic timescale of about 10 million years. SFR(UV), by contrast, also includes contributions from less massive stars and probes star formation over 100--200 million years, depending on the wavelength range, represented by the GALEX far-ultraviolet (FUV) and near-ultraviolet (NUV) bands for nearby galaxies.

Comparing the spatial distributions of SFR(UV) and SFR(\Ha) makes it possible to map the propagation of star formation through galactic discs over the past hundred million years. Examples include nearby dwarf galaxies in which ongoing star formation is concentrated along the walls and at the collision interfaces of giant gas shells, 200--1000 pc across: Holmberg~I \cite{Egorov2018MNRAS.478.3386E}, DDO53 \cite{Egorov2021MNRAS.508.2650E} and Sex~B \cite{Gerasimov2024MNRAS.529.1138G}. Another example is the episodic star formation in the gaseous accretion ring of NGC4324 \cite{Proshina2022AstL...48..139P}, where bright \Ha{} and ultraviolet regions alternate around the ring.

A key capability of Spektr-UV is imaging in both the ultraviolet and the \Ha{} line with its field camera unit. This will extend spatially resolved studies of star formation to substantially more distant galaxies. In NGC~4324, for example, at a distance of 26.2 Mpc, the GALEX spatial resolution was approximately 700 pc, comparable to the characteristic size of stellar complexes. Spektr-UV observations will substantially expand the sample of galaxies in which the propagation and triggering of star formation can be investigated. Potential targets include dwarf irregular galaxies, galaxies with accretion or collisional rings, and tadpole galaxies, which are suspected to have recently accreted low-metallicity gas clouds \cite{SanchezAlmeida2013ApJ...767...74S}.

\subsection{Prospects for coordinated fast radio burst observations and ultraviolet quasar spectroscopy to probe the intergalactic and circumgalactic media}

Ultraviolet quasar spectroscopy has been a primary tool for studying the IGM \cite{mcquinn2016} and circumgalactic medium (CGM) \cite{tumlinson2017} at $z\leq 1$ for decades. Hydrogen and metal absorption lines in quasar spectra probe the gas density distribution, metal ionization states and gas kinematics at high velocity resolution along individual sightlines. This method is, however, insensitive to hot, fully ionized gas, which theoretical estimates suggest contains a substantial fraction of the missing baryons \cite{Shull2012,degraaff2019}. Moreover, the ionization corrections required to interpret metal absorption lines introduce systematic uncertainties into estimates of CGM gas masses and pressures \cite{werk2014}.

Fast radio bursts (FRBs)---brief extragalactic radio pulses of uncertain physical origin --- provide a complementary probe of the circumgalactic and intergalactic media. The dispersion measure (DM) of each burst traces the integrated free-electron density along the path from source to observer. This integral is independent of the thermal state of the gas and is equally sensitive to relatively cool, photoionized material and hot, collisionally ionized gas \cite{mcquinn2014, macquart2020}. Yet, without independent information, the integrated DM cannot be separated into contributions from the IGM, the CGM of individual galactic haloes and the FRB host galaxy, let alone associated with specific gas ionization phases \cite{khrykin2024b, Connor2025}.

Joint analyses of FRB--quasar pairs with small angular separations can address these limitations by combining the complementary strengths of the two probes. Quasar absorption lines identify individual structures along the sightline, providing precise redshifts and neutral-gas column densities. This information allows the contribution of each structure to the integrated DM of the neighbouring FRB to be assessed. Conversely, the total DM directly constrains the free-electron column density, which, combined with the neutral-gas column density inferred from the quasar spectrum, can yield the ionized gas fraction without model-dependent corrections. FRB--quasar pairs could thus provide the first direct measurements of baryon content and mass resolved by ionization phase in the diffuse Universe, along individual sightlines rather than as global statistical averages \cite{khrykin2024b}.

Until recently, this approach was limited by the scarcity of precisely localized ($\leq\!1''$) FRBs. The next decade is expected to transform the available sample: next-generation radio interferometers, including DSA-2000 \cite{hallinan2019}, CHORD \cite{vanderlinde2019} and SKA \cite{macquart2015}, are expected to detect and localize hundreds of thousands of FRBs each year. Forecasts by \cite{werk2026} suggest that the number of localized bursts at $z \leq 1$ will reach $\sim 100~000$ by 2035, yielding thousands of FRB--quasar pairs with angular separations $\Theta <\!10'$, including approximately one hundred with $\Theta <\!1'$. This sample could provide the statistical basis for systematic three-dimensional mapping of the CGM and IGM. By then, the Vera C. Rubin Observatory's Legacy Survey of Space and Time (LSST) photometric survey will also have expanded quasar catalogues, further increasing the number of available pairs.

The spectrographs aboard Spektr-UF, covering the far- and near-ultraviolet range ($\sim 1150-3200$\AA), will provide access to key diagnostic absorption lines for studies of the IGM and CGM at $z < 1$. A major advantage is the high spectral resolution of the UVES echelle spectrograph ($R\sim 50,\!000$), which exceeds that of HST/COS ($R\sim 13,\!000-24,\!000$) and will better resolve narrow kinematic components in the CGM. The smaller primary mirror of Spektr-UF provides less collecting area than HST, limiting the number of faint quasars accessible in practice. Nevertheless, the closest and brightest FRB--quasar pairs will form a particularly valuable sample, enabling Spektr-UF to probe the baryon distribution in the IGM and CGM at $z < 1$ in detail.

\subsection{The contribution of ultraviolet astronomy to studies of chemical evolution}

Heavy elements in the ISM provide the building blocks of planets and dust, regulate the gas heating and cooling associated with star formation, and shape galaxy evolution. Their partition between the gas and solid (dust) phases on small spatial scales is a key input to star formation models: the abundances of heavy elements and dust directly control gas heating and cooling, and hence the ability of gas to fragment and collapse into stars. The primordial gas left by the Big Bang contained only trace amounts of elements heavier than helium and lithium. The production and dispersal of heavy elements are governed by the galactic baryon cycle described above. Intergalactic gas of nearly primordial composition accretes onto galaxies, reaches their inner regions and fuels star formation. Stars subsequently synthesize heavy elements in their interiors and, towards the ends of their lives, return them to the surrounding gas, for example through supernova explosions. These elements mix with the ISM, enriching both the gas and subsequent generations of stars. Some are transported beyond the ISM into the CGM and even the IGM, potentially affecting subsequent gas accretion. Heavy elements thus both shape and trace galaxy evolution across a wide range of spatial and temporal scales: from planet-forming regions to intergalactic gas, and from gas circulation in present-day galaxies to the early Universe, where the first stellar generations produced the first heavy elements.

The origin of the chemical elements is a central problem in astrophysics. Interest in this question now extends well beyond the field, driven in large part by its connection to the emergence of life, and gives it broader methodological significance.

Research over the past quarter-century has transformed our understanding of where and how chemical elements form, how they spread from their sources through the ISM, and how chemical evolution differs between the early and present-day Universe. It has also clarified the extent to which chemical composition can indicate an object's age or trace an evolutionary sequence. Observations, particularly those obtained in recent years, reveal substantial scatter in both heavy-element abundances and relative abundance ratios. Metallicity ($Z$ -- the mass fraction of heavy elements) can therefore serve as an age indicator for stars, galaxies and their subsystems only in a statistical sense. Current perspectives on cosmic chemical evolution are reviewed in \cite{Shchekinov2023}.

Over the past decade, major advances in reconstructing the chemical history of the Universe---and thus its history of stellar nucleosynthesis---have come from observations of ultraviolet heavy-element spectral features shifted into the optical and infrared by cosmological redshift $z$. These observations probe the emergence of the first stars and the transition from the cosmic dark ages to reionization. A quarter-century ago, this transition was associated with redshifts $z\simlt 6$, when the Universe was 940 million years old and the first metals were then thought to have appeared \cite{Songaila2001,Songaila2004,Savaglio2008}. This picture was shaped in part by observational accessibility: the strong ultraviolet CIV lines at $1548.2$\AA\, and $1550$\AA\, and SiIV lines at $1402.77$\AA\, and $1393.76$\AA\,, arising in intergalactic gas at $z=1.5-5.5$, fall within the optical range $\lambda\lambda=3000-9000$\AA\,. The sensitivity and spectral resolution available at Keck, and subsequently at the VLT, were sufficient to detect them \cite{Dodorico2013,Codoreanu2018}.

Our understanding of cosmic chemical history has since changed substantially, with consequences for the inferred emergence of the first stars and the broader picture of cosmic structure formation. The onset of star and galaxy formation has shifted to earlier times by more than half a billion years. Galaxies detected at $z\simgt 12$, when the Universe was only 370 million years old, are now widely discussed. Remarkably, many of their properties resemble those of nearby galaxies. Their chemical compositions, in particular, are often close to that of the young stellar population of the Milky Way, to which the Sun belongs. Observations of these distant regions have relied on infrared facilities \cite{Oesch2016,Jiang2021,Harikane2022,Harikane2022b}, including HST, Subaru, and the Spitzer Space Telescope, and over the past two years, the James Webb Space Telescope (JWST; see, for example, \cite{Ferrara2022,Finkelstein2022,Furtak2022,Tacchella2022}). Infrared observations are essential because young, actively star-forming galaxies emit a substantial fraction of their energy in the ultraviolet at $\lambda\simeq 1200-3000$\AA \cite{Hayward2015,Leja2019}. For galaxies at $z>10$, this radiation is redshifted into the near-infrared.

Several properties of galaxies in the pre-reionization era ($z>10$) may be reproduced to some extent in galactic nuclei. These regions can serve as analogs of compact systems with high stellar densities and correspondingly high energy densities in the surrounding ISM. Potential examples include the central regions of starburst galaxies such as M82, NGC3079, NGC4388 and NGC1068, together with ultraluminous infrared galaxies (ULIRGs) in the local Universe ($z \simlt 0.2$). Among the latter, the so-called warm ULIRGs are particularly notable for their rich ultraviolet spectra, originating from multiple bright, discrete knots in their nuclear regions \cite{Farrah2005}.

\subsubsection{Chemical elements in intergalactic flows}

Heavy elements detected in the early Universe at $z\sim 3-5.5$ were associated with absorption in the Lyman-$\alpha$ forest. These absorption lines are thought to arise in extended regions of diffuse IGM spanning $0.5-1$ Mpc \cite{Songaila2001,Songaila2004,Simcoe2006,Savaglio2008,Dodorico2013,Codoreanu2018}. A detailed analysis of one such region ($z=2.73$, when the Universe was $\simeq 2.5$ billion years old) identified numerous distinct substructures: clouds smaller than $< 1$ kpc, with overdensities $\rho/\bar\rho\sim 1000$ and absorption lines from multiple elements (H, C, O, N, Si, Al, Fe). Their metallicities were unexpectedly high, $Z\simeq 0.1-0.3Z_\odot$. The clouds lie at distances ranging from 200 to 200 kpc from galaxies that could have supplied the heavy elements \cite{Simcoe2006}. In current terminology, gas within $r\simlt 200 - 300$ kpc of galaxy centers is classified as CGM. This boundary is approximate, marking a qualitative transition from gas physically associated with the nearest galaxy to the IGM.

The regions containing these dense substructures occupy volumes set by the volume available per galaxy, $\simgt 1$~Mpc$^3$, substantially exceeding the volumes of circumgalactic gas. Crucially, enriched gas expelled from galaxies rapidly fragments into dense condensations with small geometric cross-sections and metallicities that increase towards smaller sizes, $Z \propto R^{-\alpha}$, $\alpha \simgt 1$ \cite{Dedikov2004,Schaye2007,Vasiliev2009}. Consequently, a substantial fraction of the material in these condensations escapes detection in absorption, contributing to the missing baryon and metal budgets. A promising approach is therefore to measure mass fluxes in outflows from galaxies with well-developed winds at the largest accessible distances, constraining their asymptotic behaviour within circumgalactic haloes and beyond, into the IGM. Such measurements have recently been made at optical and radio wavelengths for galaxies with winds in the local Universe \cite{Leroy2015,Bolatto2021,Hagedorn2026,Lin2026}. The spatial distribution of material expelled by starburst activity follows vertical profiles ranging from $\sim r^{-4}$ to $\sim r^{-2}$ and extends to $\sim 7 - 10$ kpc. These findings support the prospects for ultraviolet measurements of outflow profiles in galaxies near the peak of cosmic star formation at $z\sim 1-2$. Examples of such outflows are reported in several recent studies (see \cite{Noterdaeme2021,Noterdaeme2023,Coloma2023,Liu2025} and references therein). Mapping the large-scale distribution of heavy-element abundances at distances $\simgt 100 - 300$ kpc, comparable to or exceeding the sizes of circumgalactic haloes, would be of fundamental importance. The expected column densities of multiply ionized species are comparable to those observed in the Lyman-$\alpha$ forest and the warm--hot IGM: $N\sim 10^{12} - 10^{13}$ cm$^{-2}$.

\subsubsection{Chemical evolution in cosmic voids}

Measuring the chemical composition of cosmic voids is essential to understanding cosmic chemical evolution, the transport of heavy elements through extremely low-density environments and the chemical homogeneity of the Universe \cite{Stocke2007}. Over the past decade, studies of galaxies in nearby voids have revealed substantially less nucleosynthetic enrichment than in galaxies of the Local Supercluster \cite{Kniazev2018,Pustilnik2019,Pustilnik2024,Pustilnik2026}. The origin of this difference remains unclear. Environmental effects on the dynamics and evolution of individual void galaxies offer an intuitively compelling explanation, but the observational evidence is mixed. A comparison of dwarf galaxies in voids and denser environments in the nearby Universe ($z\simlt 0.03$) found no significant differences in their chemical evolution \cite{Douglass2017}. More recent work, using larger samples of dwarf galaxies in voids (353 galaxies) and filaments (311 galaxies) over a broader redshift range, $0.005<z\simeq 0.22$, found a substantial difference in the slopes of the mass--metallicity relation: $\beta\simeq 0.28$ in voids and $\beta\simeq 0.17$ in filaments \cite{Bidaran2026}. In this context, simulations for the first {\it Euclid} data release, DR1, targeting dwarf galaxies in voids at $0.8 < z <1.7$, offer promising prospects (see the detailed analysis in \cite{Euclid2026a}). Preliminary results indicate that environmental effects are particularly evident in the central regions of voids ($r\simlt 0.6 R_v$, where $R_v$ is the void size), affecting galaxy sizes and star formation rates. Of particular relevance to chemical evolution, void galaxies assemble $\sim 0.5 - 1$ billion years later than galaxies in filaments. These differences become apparent only at $z<1.15$ \cite{Euclid2026b}.

Studies of interstellar gas metallicity in galaxies across different environments reveal a complex dependence on overdensity, $Z(\delta)$, where $\delta=\rho/\langle\rho\rangle -1$. At $z<1.5$, the ISM of galaxies in low-density environments, $\delta<1$, is less enriched than that of galaxies in low-density regions \cite{Darvish2015,Chartab2021}. At higher redshifts, $z\sim 2.3$, this relation reverses \cite{Chartab2021}. These findings remain unexplained. One hypothesis links the change in $Z(\delta)$ across $z\sim 1.5$ to the exchange of enriched material between galaxies and the surrounding circumgalactic and intergalactic gas. Enhanced radiative cooling of enriched gas in dense environments is central to this interpretation. Further vacuum-ultraviolet observations at $z\sim 0 - 1.5$ are needed to establish and understand the emerging differences in chemical evolution between voids and filaments.

Measurements of the spatial distribution of heavy elements and the velocity fields traced by their ions could therefore provide important constraints on the transport and mixing of nucleosynthetic products throughout the Universe. Moreover, because voids may contain a substantial fraction of the cosmic baryon budget, measuring their matter content is important for resolving the missing baryon and metal problems. A potentially favourable circumstance is that slow cooling in the diffuse void environment may allow enriched regions to occupy large volumes and present large geometric cross-sections, making them comparatively easy to detect. Recent observations of the chemical composition of low-density intergalactic regions support this expectation: an average of approximately five OVI $1032/1038$\AA\, absorption systems is detected per quasar \cite{Mishra2025}.

Absorption lines of multiply ionized species may have equivalent widths of order $W\sim 1-10$~m\AA\, for ion column densities $N\sim 10^{12}-10^{13}$~cm$^{-2}$, as expected in voids with relative densities $\delta =\rho/\langle\rho\rangle \sim 0.1 - 0.03$. These estimates span a wide range, depending on the metallicities of clouds expelled into the IGM by active galaxies, their velocities, ram pressures, sizes and geometric cross-sections, and the spectrum of the ionizing ultraviolet radiation. Preliminary estimates based on observations of warm--hot gas in the local IGM and CGM, together with constraints on multiply ionized species at $z = 4 -6$, yield $N \sim 10^{11} - 10^{14}$~cm$^{-2}$ and covering fractions $f\sim 0.003 - 0.1$, depending on the physical conditions and morphology of the condensations. Values close to these approximate estimates, $N({\rm OVI})\sim 10^{12}$~cm$^{-2}$ and $W_{1032\text{\AA},}\approx 1.7$~m\AA\, were recently obtained from measurements of 396 OVI doublet absorption lines at $z\sim 0.1 - 0.5$ in the spectra of 82 quasars~\cite{Mishra2025}.

\subsection{The missing baryon problem}

The missing baryon problem concerns the apparent deficit of baryonic matter in the present-day Universe relative to the abundance measured at high redshift and predicted by cosmological models. Baryons contained in galaxies account for only about 10\% of the abundance, $\Omega_{b}=0.045$, inferred from primordial nucleosynthesis, cosmic microwave background (CMB) measurements and Ly-$\alpha$ forest absorption \cite{Shull2003,Fukugita2004,Shull2012}; see also the review by \cite{Bregman2007}. Most baryons appear to reside in intergalactic gas. A recent analysis combining fast radio burst dispersion measurements with gravitational-wave observations yielded $\Omega_b=0.0488\pm 0.0064$ \cite{Zhang2025}. This agrees closely with the value $\Omega_b=0.0488\pm 0.0064$ inferred from Planck CMB data. A similar estimate, $\Omega_b=0.051\pm 0.006$, was obtained from DMs in an expanded sample of FRBs with measured redshifts $z>1$ \cite{Connor2025}. Direct baryon inventories in the local Universe, however, remain substantially below these values: optical, ultraviolet and X-ray observations of galaxies, galactic haloes and warm--hot gas still leave approximately 60\% of the baryon budget unaccounted for \cite{Dodorico2023}.

The warm--hot intergalactic medium (WHIM) is thought to be a major reservoir of missing baryons \cite{Cen1999,Cen2006}; see also the reviews in \cite{Prochaska2009,Peroux2020}. Recent dispersion measurements of distant FRBs at $z>1$ support a predominantly intergalactic baryon reservoir, placing 76\% of baryons in the IGM. Despite extensive efforts in recent years, however, direct measurements of the baryonic mass in the WHIM remain limited. One obstacle is the strong observational selection associated with the slow mixing of heavy elements, whose ions provide the principal tracers of the WHIM \cite{Dedikov2004,Vasiliev2009,Shchekinov2022,Shchekinov2023}. Further uncertainties arise from the ionization state and phase structure of the gas \cite{Simcoe2006,Goetz2024,Koplitz2024}. A key limitation is the inability to separate absorption and emission contributions from distinct regions at the spectral and angular resolutions of existing instruments. Under these conditions, hybrid approaches that combine observations with numerical simulations to constrain models offer a practical way forward (see, for example, \cite{Simcoe2006}).

\subsection{Tidal flows and galactic tails}\label{sec:shchekinov}

Large-scale flows generated by tidal interactions between galaxies provide a valuable laboratory for studying the transport and mixing of heavy elements in astrophysical environments. Metal mixing by tidal flows during galaxy mergers was first described in \cite{Gnedin1998}. Ram-pressure stripping of galactic gas haloes by the surrounding medium provides another efficient mechanism for redistributing metals and promoting their subsequent mixing \cite{Dedikov2004}. This mechanism is likely to dominate in galaxy clusters, where gas densities are substantially higher than in the IGM. Mixing in enriched gas is nevertheless hindered by the development of density inhomogeneities, driven by the density dependence of radiative cooling. These inhomogeneities lengthen the characteristic timescales of the processes responsible for mixing. As a result, the intracluster gas exhibits spatial variations in metallicity even in virialized clusters \cite{Lovisari2011,Mernier2018}. Understanding the physical origin of inefficient mixing requires a detailed analysis of the kinematics of a multicomponent fluid, as appropriate for intracluster gas with an inhomogeneous chemical composition. Extended tails of cluster galaxies offer an opportunity for such an analysis: shear flows at their outer boundaries drive Kelvin--Helmholtz instabilities, providing a classical setting for fluid mixing.

Recent studies have documented numerous galaxies with bright tails ($f({\rm H\alpha +[NII]})\sim 10^{-11} - 10^{-15}$ erg cm$^{-2}$ s$^{-1}$) extending over $\sim 200$ kpc (see the review by \cite{Boselli2022}). Gas densities and temperatures in these tails span wide ranges, $n \sim 10^{-2} - 10^{-5}$ cm$^{-3}$ and $T\sim 10 - 10^7$ K, with an approximately inverse relation, $T\sim n^{-1}$ \cite{Boselli2022}. These galaxies belong to the nearby Virgo and Coma clusters, so the gas associated with their tails spans angular regions of $\simgt 1^\circ \times 1^\circ$. The expected angular resolution of Spektr-UF will enable detailed studies of the spatial distribution and velocity fields of chemical species, together with the physical state of the gas.

\section{The interstellar medium of the Milky Way and molecular clouds}

\subsection{Ultraviolet observations of diffuse neutral gas}\label{sec:diffuse_ISM}

Diffuse neutral gas in the ISM is the main reservoir of material available for star formation and contains most of the metals during the first half of cosmic history \cite{Peroux2020}. Its thermal balance gives rise to two phases: a relatively dense cold neutral medium at $T\sim 50-100$\,K and a diffuse warm neutral medium at approximately $5000-8000$~K \cite{Field1969,Wolfire2003}. A fundamental observational challenge is its extremely low emissivity, which makes direct emission-line observations impractical across most spectral bands. Absorption spectroscopy is therefore the principal tool for studying this gas.

\begin{figure*}
	\includegraphics[width=0.95\linewidth]{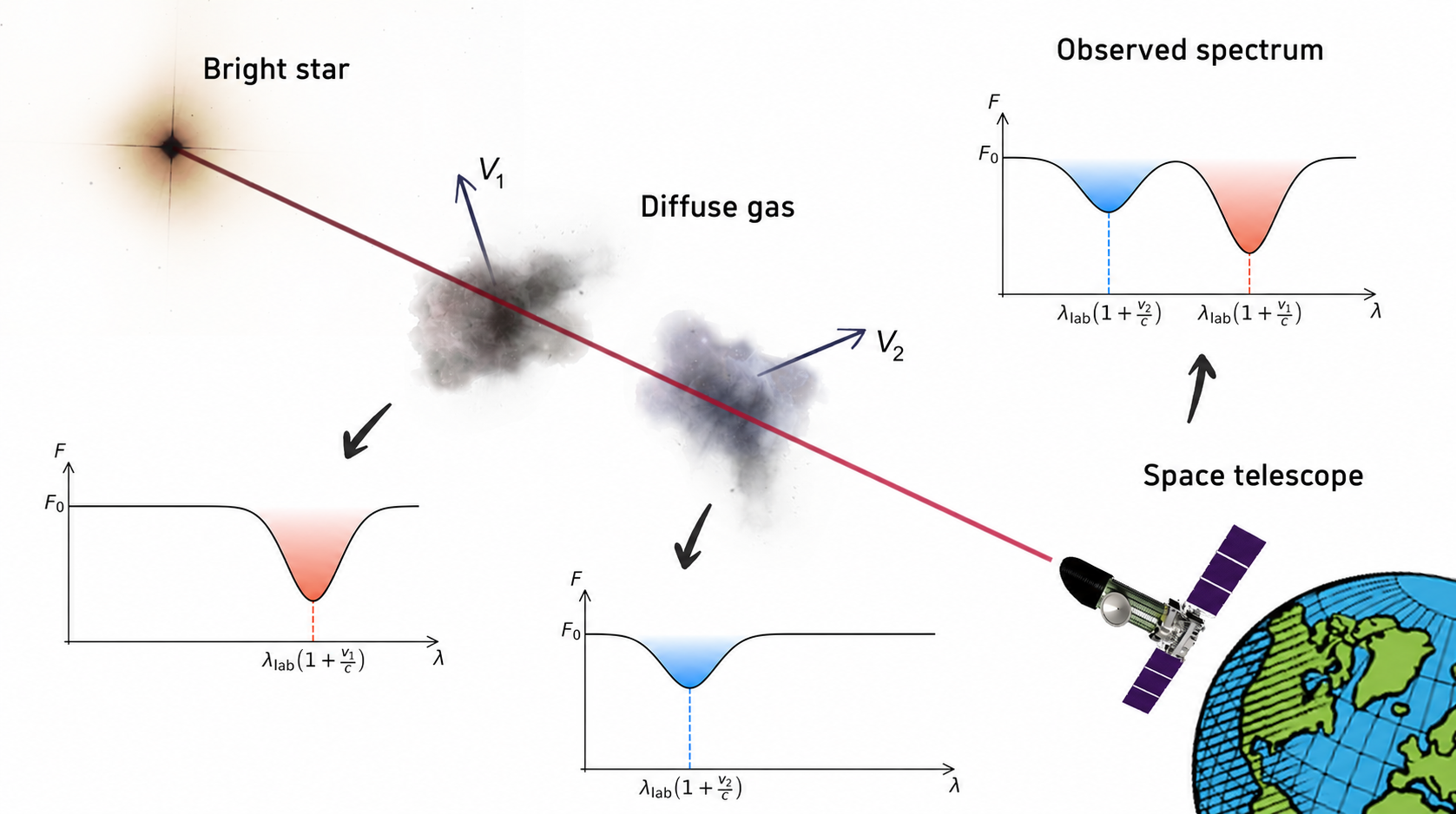}
	\caption{Principles of ultraviolet absorption spectroscopy for studies of the diffuse ISM. A space-based telescope records the ultraviolet spectrum of a bright background star, revealing absorption lines produced by intervening clouds. Clouds in different regions can have distinct line-of-sight bulk velocities, allowing their absorption components to be separated spectroscopically and elemental abundances to be measured in individual diffuse clouds.}
	\label{fig:abs_principles}
\end{figure*}

Ultraviolet absorption lines constrain the physical conditions \cite{Jenkins2001,Balashev2019,Kosenko2024} and gas-phase abundances of heavy elements in the ISM. Relative elemental abundances also provide indirect constraints on the dust content and composition \cite{Field1974,Savage1996,Jenkins2009,DeCia2016}. Figure~\ref{fig:abs_principles} illustrates the principles of ultraviolet absorption spectroscopy.

The ultraviolet contains the highest density of atomic and molecular resonance lines, tracing a wide range of ISM phases. These transitions arise predominantly from the ground states of atoms, ions and molecules. Absorption spectroscopy therefore allows gas-phase elemental abundances to be measured directly, without the level-population corrections and complex modelling required to interpret emission lines. Because the background stars are point sources, absorption spectra are essentially free from the angular averaging inherent in emission observations. Moreover, separating absorption components by their line-of-sight bulk velocities enables detailed measurements of elemental abundances and physical conditions in individual diffuse interstellar clouds.

There is no universally accepted definition of diffuse clouds \cite{2017ASSL..442.....M}, although the classification introduced in \cite{1988ApJ...334..771V} is widely used. In this scheme, clouds with $A_{\rm V}<1^{\rm m}$ are classified as diffuse, whereas those with $1^{\rm m}<A_{\rm V}<5^{\rm m}$ are termed translucent. These categories are particularly relevant to observations with Spektr-UF, because clouds with greater extinction are opaque to ultraviolet radiation. Strictly speaking, diffuse and translucent clouds need not be discrete objects: they may instead form the low-density envelopes of denser, dark clouds. For simplicity, we use the term diffuse clouds below to include both diffuse and translucent material.

\subsection{Molecules in diffuse gas}\label{sec:wiebe}

Diffuse molecular clouds typically have number densities of a few hundred~cm$^{-3}$ and temperatures of about 100~K or lower. Hydrogen is not yet fully molecular, and carbon is predominantly ionized or atomic, indicating exposure to ultraviolet radiation. Nevertheless, these environments harbour a rich molecular inventory \cite{2006ARA&A..44..367S}. Their chemistry is of considerable interest because the processes operating under these conditions remain poorly understood, even setting aside the problem of the diffuse interstellar bands. Interstellar molecules were first discovered in diffuse gas, yet the origin of the first detected molecular ion, CH$^+$ \cite{1941ApJ....94..381D}, remains a puzzle (see, for example, \cite{2023A&A...669A..74G}). More complex molecules, including formaldehyde and other organic compounds, have since been identified in diffuse environments \cite{2015MNRAS.452.3969C,2024A&A...686A..49G}.

Understanding diffuse gas chemistry and its connection to the chemistry of molecular clouds requires observations of more molecular species along more sightlines, spanning a broader range of physical conditions. Ultraviolet molecular absorption spectroscopy has played, and continues to play, a central role in this effort.

Molecular absorption in the ultraviolet is both a diagnostic tool and an important driver of chemistry. Overlap between the absorption bands of H$_2$ and CO allows the more abundant H$_2$ to shield CO from ultraviolet radiation and hence photodissociation~\cite{vanDishoeck_88}. This shielding couples the abundances of H$_2$ and CO, making CO emission a widely used tracer of molecular hydrogen in the ISM. Their Galactic column densities are commonly assumed to satisfy $N({\rm H_2}) \approx N({\rm CO})\times 10^4$, with the precise conversion factor set by the local elemental carbon abundance. However, observations of CO, CH and CH$^+$ absorption in diffuse clouds~\cite{Federman_1982, Sheffer_2008} show that this relation breaks down at $N({\rm H_2}) < 10^{21}$~cm$^{-2}$. CO is then orders of magnitude less abundant than expected, whereas the CH radical becomes a tracer of H$_2$.

Two distinct chemical regimes can be identified in diffuse clouds. At $N({\rm CO}) > 10^{14}$~cm$^{-2}$, gas-phase CO is maintained through the reaction $\rm C^+ + OH \rightarrow CO^+ + H$, followed by $\rm CO^+ + H \rightarrow CO + H^+$. At lower CO column densities, the dominant pathway begins with $\rm CH^+ + O \rightarrow CO^+ + H$, followed by $\rm CO^+ + H_2 \rightarrow HCO^+ + H$ and recombination through $\rm HCO^+ + e^- \rightarrow CO + H$. Reproducing the observed CO abundances in this regime requires chemical models that go beyond equilibrium assumptions. One possibility is to include the dissipation of Alfv\'en waves in diffuse clouds \cite{Federman_1996, Zsargo_2003}, which enhances CH$^+$ formation.

Observations with Spektr-UF of absorption by CO, CH, CH$^+$ and other molecules characteristic of diffuse gas towards ionizing stars could clarify CO photodissociation and the chemical links between these species. Ultraviolet radiation from young stars vibrationally excites H$_2$, enabling endothermic reactions that are suppressed in the shielded interiors of molecular clouds \cite{Agundez_2010}. The relationships between the column densities of CO, CH, CH$^+$ and H$_2$ are therefore expected to differ between diffuse clouds near young stars and the dense photodissociation regions studied with infrared spectroscopy (for example, \cite{2024A&A...685A..73H}). Spektr-UF measurements of CO, CH and CH$^+$ absorption will constrain their column densities in diffuse clouds and inform the development of non-equilibrium chemical models.

Water is of particular interest in studies of diffuse interstellar chemistry. Whereas ultraviolet spectra of T Tauri stars probe water abundances under extreme temperatures and radiation fields, searches for water absorption in diffuse gas can illuminate its formation under cold conditions. Several attempts to detect ultraviolet water absorption towards nearby bright stars were unsuccessful \cite{1998ApJ...503..780S}, although water has been detected at submillimetre wavelengths \cite{2021A&A...648A..24V}. This contrast may reflect the temperature dependence of the chemical pathways that produce water. The sensitive spectrograph aboard Spektr-UF will enable this possibility to be investigated.

\subsection{Selective photodissociation}

In addition to the main CO isotopologue, rarer species such as C$^{18}$O, C$^{17}$O and $^{13}$CO are also observed. Their ionization and dissociation cross-sections differ \cite{refId0}, causing isotopologue abundance ratios to depart from the corresponding elemental isotope ratios at molecular cloud edges, where ultraviolet radiation is only partially attenuated and overlap between absorption lines affects shielding. An analysis of \hh, CO and $^{13}$CO column densities measured with HST was presented in \cite{2007ApJS..168...58S}. Plots of $N({\rm CO})/N({\rm H}_2)$ and $N(^{13}{\rm CO})/N({\rm H}_2)$ against $N({\rm H}_2)$ reveal the onset of CO shielding by H$_2$. However, insufficient observational coverage prevented a comparable assessment for $^{13}$CO. High-quality Spektr-UF spectra containing absorption lines of $^{13}$CO, C$^{18}$O and C$^{17}$O could fill this gap and quantify the effects of differential shielding and selective photodissociation on the rarer CO isotopologues. Selective shielding has, for example, been detected in molecular gas towards $\rho$~Oph~A and $\chi$~Oph \cite{2003ApJ...591..986F}. Beyond constraining ultraviolet shielding, observations of $^{13}$CO, C$^{18}$O and C$^{17}$O can also probe the efficiency of isotope-exchange reactions in cold, dense gas. Enhanced $^{13}$CO abundances relative to $^{12}$CO were reported in \cite{2007ApJS..168...58S} along sightlines where the $^{12}{\rm CO}/^{13}$CO ratio falls below the mean Galactic $^{12}$C/$^{13}$C ratio.

\subsection{The reddening law in photodissociation regions} \label{sec:kirsanova}

The surroundings of young massive stars still embedded in their natal molecular clouds exhibit a remarkable diversity of ionization states, physical conditions and velocity fields. Observations spanning radio to ultraviolet wavelengths reveal this complexity. Radio, millimetre and far-infrared observations trace dense molecular filaments and clouds; optical observations reveal ionized nebulae occupying the intervening space; and near- and mid-infrared observations probe photodissociation regions (PDRs), where ultraviolet radiation drives molecular photodissociation, particularly of \hh. Carbon, sulphur and silicon are ionized in PDRs. Their relatively high densities and temperatures support active chemistry, producing a rich molecular inventory. Molecular shielding also establishes photodissociation fronts, allowing substantial molecular populations to survive under distinctive physical conditions. The nearby Orion massive star-forming region and the HII region surrounding the Trapezium stars have been studied extensively (for example, \cite{2006ApJ...644..344A}). Most other HII regions, however, remain only partially characterized, despite the wide range of spectral types of their ionizing stars and nebular morphologies.

Molecular emission lines offer diagnostics across the full range of PDR conditions, from the photodissociation front to the boundary with shielded molecular gas. They provide powerful probes of the evolution of molecular clouds exposed to radiation from massive stars. The wavelength coverage of the Spektr-UF spectrographs enables observations of ultraviolet molecular emission, particularly from \hh{} and CO at 1200--1700\AA. Such observations are needed to resolve PDR structure, constrain the thermal balance of the ISM and investigate the microscopic processes underlying its diverse properties. Below, we discuss results from molecular-line observations of diffuse clouds that may also apply to PDRs, but require verification through observations with Spektr-UF.

Ultraviolet spectroscopy with Copernicus established an empirical relation between the hydrogen column density and optical reddening. An analysis of Ly-$\alpha$ absorption towards bright stars \cite{1978ApJ...224..132B} yielded a mean ratio $<N(\rm{HI+H_2})/E(B-V)> = 5.8\cdot 10^{21}$~hydrogen nuclei cm$^{-2}$~mag$^{-1}$, where $ N(\rm{HI+H_2}) = N(\rm{HI})+ 2N(\rm{H_2}) $. These measurements were restricted to stars with $E(B-V) \leq 0.5$. Adopting a total-to-selective extinction ratio $R_V=3.1$ gives $N(\rm{HI+H_2})=A_V\cdot 1.87\cdot10^{21}$~cm$^{-2}$, a relation widely used to estimate hydrogen column densities from extinction measurements. Subsequent FUSE spectroscopy \cite{2002ApJ...577..221R, 2009ApJS..180..125R} yielded a similar relation, $N(\rm{HI+H_2})=A_V\cdot 1.9\cdot10^{21}$~cm$^{-2}$, for sightlines towards stars with $A_V\leq5$~mag, again assuming $R_V=3.1$ in the molecular cloud. These studies also found that the molecular contribution to $N(\rm{HI+H_2})$ decreases with increasing $R_V$, although the scatter in the \hh{} fraction increases. This scatter warrants further observations and explanation, particularly because larger dust grains are expected to reduce the rate of molecular hydrogen formation.

In HII regions and PDRs, a deficit of small dust grains modifies the reddening law, producing higher values of $R_V$, up to 5.5. Examples include the Orion Bar \cite{1998A&A...330..696M}, NGC~7023~\cite{2006ApJ...636..303W} and the PDRs surrounding the HII regions S255-257~\cite{2023MNRAS.526.5187K}. Observations of H$_2$ or Ly-$\alpha$ absorption towards the ionizing stars of HII regions could establish how the empirical Copernicus and FUSE relations change in environments containing larger grains. Although $A_V$ in HII regions can be estimated from ground-based optical observations of the H-$\alpha$ and H-$\beta$ recombination lines, measurements of \hh{} or Ly-$\alpha$ absorption require an ultraviolet telescope. Transitions involving the ground and/or low-lying states of \hh{} are inaccessible to Spektr-UF because the far-ultraviolet coverage of UVES and the long-slit spectrograph begins at 1150~\AA~\cite{Shugarov_2023}. However, transitions involving vibrationally excited \hh{} can be observed \cite{1993A&AS..101..273A, 1993A&AS..101..323A}, allowing the \hh{} column density to be estimated by modelling the physical conditions in the PDR. An example of this approach using HST observations is presented in \cite{2003ApJ...591..986F}.

\subsection{Spectral formation in HII regions and PDRs} \label{sec:kirsanova2}

\begin{figure*}
 \includegraphics[width=0.95\linewidth]{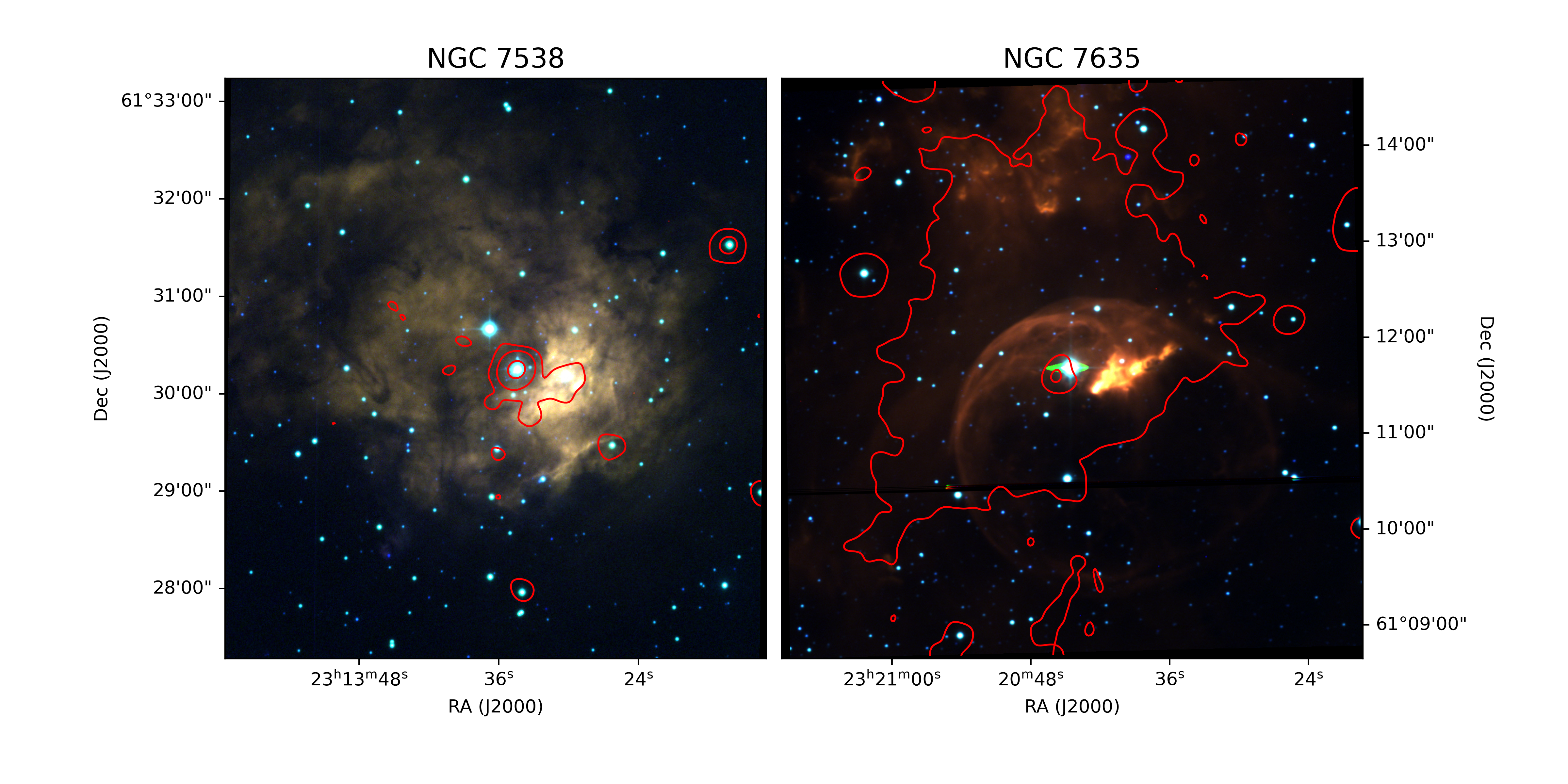}
 \caption{Ultraviolet emission contours from GALEX~\cite{2007ApJS..173..682M} (red), overlaid on optical images of NGC~7538 and NGC~7635 constructed from archival IPHAS observations in the r, i and Ha filters \cite{2005MNRAS.362..753D}. For NGC~7538, the contours correspond to $S/N = 2, 4, 45$, estimated from the GALEX~NUV image after $5\times5$ binning. For NGC~7635, the contours correspond to $S/N = 2, 255, 364$, estimated from the GALEX~NUV image after $4\times4$ binning.}
 \label{fig:galex_iphas}
\end{figure*}

Ultraviolet spectra of HII regions reveal fluorescent \hh{} emission, apparently driven by radiative pumping by photons from the ionizing stars, which populates highly excited rovibrational states \cite{2005ApJ...628..750F, 2005ApJ...629L..97F}. Obtaining ultraviolet spectra along sightlines with existing infrared observations of \hh{} emission would be particularly valuable. The infrared lines trace the relaxation of molecules towards the vibrational ground state. Together, these complementary datasets could constrain the contributions of radiative pumping, shocks and collisions with non-thermal electrons to the \hh{} emission spectrum. Suitable targets include nebulae in the OPTIMus multiwavelength survey of HII regions \cite{OPTIMUS}, for which extensive \hh{} observations are available \cite{2023AstBu..78..372K, 2026arXiv260414863G}. Examples of OPTIMus nebulae are shown in Fig.~\ref{fig:galex_iphas}.

A study of IC~63 \cite{2005ApJ...628..750F} reported several lines associated with excited ionic levels, including CII$^*$, NII$^*$ and SIII$^{**}$, whereas transitions to the ground state were either undetected or unexpectedly weak. The authors attributed this behaviour to self-absorption within the nebula and suggested that ionized gas extends beyond its optically visible boundaries. Extending such measurements with Spektr-UF would enable detailed studies of ionic excitation and the structure of HII regions.

As noted above, the wavelength coverage of the VUVES spectrograph planned for Spektr-UF begins at $\sim$1150~\AA. Observations in the local Universe will therefore access H$_2$ transitions involving highly excited states rather than the principal ground-state lines. Analysis of H$_2$ level populations constrains the gas density, temperature and ultraviolet radiation field in PDRs. Moreover, observations of numerous absorption lines arising from different levels can reveal subtle radiative-transfer effects (for example, \cite{Lacour2005,Balashev2009}) and constrain the hydrogen ortho-to-para ratio, another sensitive diagnostic of physical conditions and H$_2$ formation \cite{Fuente1999,SternbergNeufeld1999}.

\subsection{Elemental abundances in the local Universe}
\label{sec:metals_ISM}

How heavy elements mix within galaxies and their surroundings is central to understanding galaxy formation and chemical evolution \cite{Matteucci2021}, yet remains largely unresolved. Numerical simulations predict substantial metallicity variations in the circumgalactic medium, driven by gas inflows (for example, \cite{LehnerHowk2011,FoxDave2017}, with accretion rates of $\approx$0.1–1 $M_{\odot}$~yr$^{-1}$) and outflows (for example, \cite{Peroux2020}). Observations of galaxies provide support for this picture (for example, \cite{Cameron2021,Wendt2021}). Conversely, Galactic rotation could, in principle, mix the ISM over a volume of approximately 0.1~kpc$^3$ within a few $10^8$ years \cite{Edmunds1975}. How long low-metallicity gas can survive without mixing into the surrounding ISM has implications for several areas of astrophysics, including the detectability of pair-instability supernovae \cite{Cikota2017} and searches for the chemical signatures of the first, Population III stars \cite{DOdorico2024}.

Measurements of B stars and HII regions in the Milky Way reveal a shallow metallicity gradient (--0.037~dex~kpc$^{-1}$) with little scatter (0.07~dex; \cite{ArellanoCordova2020}). Young disc stars, however, exhibit variations of $\sim0.5$~dex, interpreted as dilution by accreted gas \cite{Spitoni2023}. Absorption spectroscopy shows that the depletion of heavy elements into dust in the neutral Galactic ISM varies substantially between sightlines \cite{Jenkins2009} and along individual sightlines \cite{Welty2020}. Recent studies have also reported substantial metallicity variations in the neutral ISM, reaching values as low as approximately 20\% of solar \cite{DeCia2021,RamburuthHurt2025}.

A subsequent absorption-line analysis \cite{Ritchey2023}, however, found no departures from solar metallicity larger than $\sim$0.10~dex in the neutral ISM. Notably, the large variations reported by \cite{DeCia2021} were inferred mainly from moderately refractory elements, whereas \cite{Ritchey2023} relied predominantly on volatile elements and found high metallicities with little variation. Refractory and volatile elements can weight the absorbing material differently, depending on their sensitivity to high- and low-metallicity gas. Both studies nevertheless identified anomalous abundance patterns, indicating that the ISM comprises overlapping clouds with distinct properties. High-resolution spectroscopy with Spektr-UF could provide new insights into interactions between stars and the ISM, stellar feedback, and the transport and enrichment of heavy elements. Ultraviolet observations would address two key questions:

\begin{enumerate}
\item What is the intrinsic chemical composition of the Milky Way ISM?

This question extends beyond measuring individual elemental abundances. It requires a comprehensive inventory of key elements (C, N, O, S, P, Si, Cl, Cu, Ca, Na, K, Ti, Mn, Fe, Co, Ni, Zn, Kr and others), accounting for their depletion from the gas phase into dust. A substantial fraction of some elements---up to 99\% for iron---may be locked in solid grains. Recovering the total chemical composition therefore requires precise atomic and ionic column densities from absorption lines, together with constraints on depletion. Carbon is particularly important, both as a fundamental constituent of life and as a major component of cosmic dust. Carbonaceous grains account for a substantial fraction of Milky Way dust, yet the total amount of carbon in the solid phase remains poorly constrained. The principal difficulty is that the only suitable transition, C\,II]\,2325~\AA, is exceptionally weak. High-resolution echelle spectra obtained with HST's Space Telescope Imaging Spectrograph (STIS) have yielded only five secure detections of C\,II]\,2325~\AA\ in the diffuse ISM \cite{Sofia2004}; several earlier detections with HST's former Goddard High Resolution Spectrograph (GHRS) have substantially larger uncertainties \cite{Sofia1998}. Detecting this line requires both high spectral resolution and a high signal-to-noise ratio. Even if carbon depletion is relatively modest, its high cosmic abundance means that small changes in the depletion level\footnote{The selective removal of particular constituents from a mixture. In astrophysics, depletion refers to the removal of elements from the gas phase through their incorporation into dust.} produce substantial changes in the inferred amount of carbon in dust \cite{Jenkins2009}. Similar observational limitations affect Si\,II measurements using the weak 2335~\AA\ line \cite{Jenkins2009} and oxygen measurements using O\,I 1355~\AA. The C\,II\,1334~\AA\ transition is usually fully saturated, precluding precise column-density measurements.

\item Is the ISM chemically homogeneous on small scales?

This question tests a longstanding assumption in the interpretation of observations: that the ISM near the Sun is well mixed and chemically homogeneous. A sightline through diffuse gas may intersect several clouds with distinct kinematics and enrichment histories. The data quality achievable with Spektr-UF would allow absorption profiles to be decomposed into individual kinematic components, enabling heavy-element abundances to be measured separately in each component.
\end{enumerate}

Combining Spektr-UF observations with existing catalogues of bright O- and B-type stars would enable analyses of dozens of Galactic sightlines spanning different Galactocentric distances and ISM phases. These measurements would provide a representative picture of chemical enrichment and inhomogeneity in the nearby Galactic disc. Systematic ultraviolet observations could also reveal temporal variations in the column densities of individual velocity components \cite{Price2001,Welty2007}, offering a means to probe the evolving ISM in real time.

\subsection{The balance between the cold and warm neutral media}

Spektr-UF observations could address the unresolved distribution and balance of the cold and warm neutral phases in the Galactic ISM. Theory predicts that these phases should coexist in approximate pressure equilibrium. Yet observational constraints on their relative volume filling fractions in the Galactic disc, and on how these fractions depend on local conditions, remain incomplete and conflicting. A further challenge is that ultraviolet absorption from both phases can occur along the same sightline and must be reliably separated. This can be achieved using tracers of cold gas, such as C\,I and Cl\,I, together with comparisons of gas kinematics, line widths and fine-structure excitation in C\,I, C\,II and Si\,II. A statistical analysis of high-resolution spectra along a large sample of sightlines through different regions of the disc and halo would constrain the relative contributions of the cold and warm phases. It would quantify the fraction of Galactic neutral hydrogen in a cold, dense state potentially available for star formation, and establish how this balance is regulated by local chemical composition, dust abundance and energy input from massive stars.

\subsection{Small-scale structure in interstellar clouds} \label{sec:zinchenko}

Substantial indirect evidence points to small-scale inhomogeneities in interstellar clouds. The anomalous hyperfine structure of the ammonia (1,1) line can be explained by a model in which the emission arises from numerous small ($ \sim 10^{-2} $~pc), dense ($ \sim 10^6 - 10^7 $~cm$^{-3}$) clumps with narrow intrinsic lines ($ \lesssim 0.3$~km~s$^{-1}$) \cite{Stutzki1985}. The observed hyperfine structure of HCN $ J=1-0 $ is also best reproduced by a clumpy model \cite{Pirogov1999}. An analysis of water excitation in the S140 PDR likewise indicates dense clumps embedded in a more diffuse medium \cite{Poelman2005}. Comparisons of observations of different molecular isotopologues provide further evidence for cloud fragmentation \cite{Martin1984}. Several studies have examined small-scale ripples in spectral line profiles. These ripples arise when emission is produced by an ensemble of clumps whose internal velocity dispersions are much smaller than the dispersion of their relative motions (for example, \cite{Tauber1996}). Analyses of HCN and HCO$^+$ $ J=1-0 $ and CS $ J=2-1 $ profiles in several star-forming regions detected such ripples in some sources and constrained the properties of the emitting gas \cite{Pirogov2008, Pirogov2012, Pirogov2018}. Modelling yielded characteristic clump sizes of $ \sim 40$~au and a volume filling factor $f_V \sim 0.01$ \cite{Pirogov2018}. Recent ALMA observations of the dense S255IR core at a resolution of $ \sim 20$~au directly reveal its clumpy structure \cite{Zinchenko2023}.

Interstellar inhomogeneities can also be studied in absorption against sufficiently bright, compact background sources. A millimetre-wavelength example is provided by \cite{Liszt2000}: observations of HCO$^+$ $ J=1-0 $ absorption towards several sources, separated by approximately two years, revealed substantial changes in the line profiles. The authors attributed these changes to small-scale variations in chemical composition.

Similar measurements are possible at other wavelengths, including the ultraviolet, and several examples are already available. Monitoring of optical and ultraviolet interstellar absorption towards HD\,219188 revealed changes of more than a factor of two in the Na~{\sc i} and Ca~{\sc ii} column densities of one velocity component over several years \cite{Welty2007}. Absorption by the dominant neutral or ionic species S~{\sc ii}, O~{\sc i}, Si~{\sc ii} and Fe~{\sc ii} remained nearly constant, indicating a stable total hydrogen column density, estimated at $ \sim 6\times 10^{17} $~cm$^{-2}$. Variations in C~{\sc i} and C~{\sc ii}$^*$ closely tracked those in Na~{\sc i} and Ca~{\sc ii}. These results imply changes in hydrogen number density from $ \sim 20 $~cm$^{-3}$ to $ \sim 45 $~cm$^{-3}$ and in electron density by more than a factor of three on scales of tens of au. Variations of 40\% in the Na~{\sc i} column density were also detected across $ \sim $50~au in the Local Leo Cold Cloud~\cite{Meyer2012}. Small-scale structure in the diffuse ISM is reviewed in detail by \cite{Stanimirovic2018}.

Spektr-UF could substantially extend these studies. Increasing the number of sightlines through diffuse interstellar gas would allow small-scale structure to be characterized across different environments. Similar measurements could also be attempted in denser regions, provided that they remain sufficiently transparent to ultraviolet radiation. This requirement corresponds approximately to a hydrogen column density $ N\lesssim 10^{21} $~cm$^{-2}$, given that ultraviolet extinction is several times greater than optical extinction \cite{Cardelli1989}. For clump sizes of $ \sim $20~au and gas densities of $ \sim 10^5 - 10^6 $~cm$^{-3}$, the hydrogen column density through an individual clump is of order $ \sim 10^{20} $~cm$^{-2}$. With a clump volume filling factor $f_V \sim 0.01$, the condition $ N\lesssim 10^{21} $~cm$^{-2}$ would hold over a region of size $ L\lesssim 20000 $~au, or 0.1~pc. Such cores exist, but suitable targets would also require sufficiently bright ultraviolet sources behind or within them. Young massive stars or protostars are the most likely candidates. A compact group of such sources would enable spatial fluctuations in absorption to be measured. These estimates remain uncertain; in particular, the more diffuse gas surrounding the dense clumps also contributes to ultraviolet extinction.

Suitable diagnostics include the lines observed by \cite{Welty2007}, together with other atomic and ionic transitions. Molecular absorption, particularly from H$_2$ and CO, could also be sought in dense clumps, although their strongest lines lie outside the wavelength coverage of Spektr-UF.

\subsection{Nebulae around high-velocity pulsars}\label{sec:BSPWNe}

Core-collapse supernovae produce neutron stars, including pulsars. Supernova remnants constitute a broad research field in their own right; here, we focus on nebulae formed by interactions between the ISM and relativistic pulsar winds \cite{2020MNRAS.497.2605B}. Pulsars provide distinctive probes of the density of the warm ISM.

Recent HST observations have yielded the first ultraviolet detections of several such nebulae through the broad-band F125LP and F140LP filters (1250 -- 2000~\AA{} and 1350 -- 2000~\AA{}, respectively) \cite{2016ApJ...831..129R, 2017ApJ...835..264R, 2025A&A...696A.121A}. The sources and their properties are listed in Table~\ref{table:pwne-known-sources}. The nebula around PSR J0437–4715 was also detected with the HST PR130L prism \cite{2016ApJ...831..129R}, and the apex of the PSR J1741-2054 nebula was detected through the $b_\text{high}$ filter on the VLT \cite{2025A&A...696A.121A}.

The origin of the observed emission remains uncertain. Existing vacuum-ultraviolet photometry and spectroscopy are consistent with both a power-law spectrum and emission from a plane-parallel shock propagating at $\approx 100$~km~s$^{-1}$ \cite{2016ApJ...831..129R, 2017ApJ...835..264R}.

\begin{table}
\caption{Properties of known ultraviolet-detected nebulae around high-velocity pulsars \cite{2025A&A...696A.121A}. Columns give the pulsar designation, spin-down power (assuming a moment of inertia of $10^{45}~\text{g cm}^2$), transverse pulsar velocity, \HaLine{} flux, flux and luminosity in the 1250 -- 1850~\AA{} band assuming a power-law spectrum, and the HST F125LP-to-F140LP count-rate ratio.}
\label{table:pwne-known-sources}
\begin{center}
\begin{tabular}{c|c|c|c|c|c|c}
\hline
PSR        & $\dot{E}$, & $V_\perp$, & $F_\text{\HaLine}$, & $F_\text{FUV}$, & $L_\text{FUV}$ & $F_\text{F125LP} / F_\text{F140LP}$ \\
           & $10^{33}$~erg~s$^{-1}$ & km~s$^{-1}$ & $10^{-14}~\dfrac{\text{erg}}{\text{s}~\text{cm}^2}$ & $10^{-14}~\dfrac{\text{erg}}{\text{s}~\text{cm}^2}$ & $10^{30}$~erg~s$^{-1}$ \\
\hline
J1741-2054 & 9.5 & 140 & 2.3 & 10 [>6]  & 4.7 [>2.8] & 1.6 $\pm$ 0.1 \\
J0437-4715 & 5.5 & 104 & 3.3 & 15 -- 22 & 0.7 -- 1.0 & 1.6 $\pm$ 0.1 \\
J2124-3358 & 6.8 & 101 & 1.7 & 5        & 1.3        & 1.3 $\pm$ 0.2 \\
\hline
\end{tabular}
\end{center}
\end{table}

Lines observed in supernova remnants are natural diagnostics of these pulsar nebulae, as they arise through the same mechanism: relatively slow but still strong shocks. Suitable transitions must also permit calculations of the relevant ionization states and fall within the sensitivity range of Spektr-UF. These criteria are met by \LaWL, \CIIlineShort, \NIVline, \CIVline, \HeIIlineLong, \OIIIline{} and \NIIIline, \OIIlineLong, \OIIlineShort{} and \NeIVline. Other promising diagnostics include \HeIIlineShort, which overlaps \La, together with \CIIlineLong{} and \CIIIline{}; the latter two have been observed to be bright in some explosive events.

\subsubsection{Dependence of nebular luminosity on pulsar velocity \label{subsubsec:pwne-luminosity-from-velocity}}

\begin{figure} \centering
        \includegraphics[width=0.8\linewidth, angle=-0, trim=0 0 0 0]{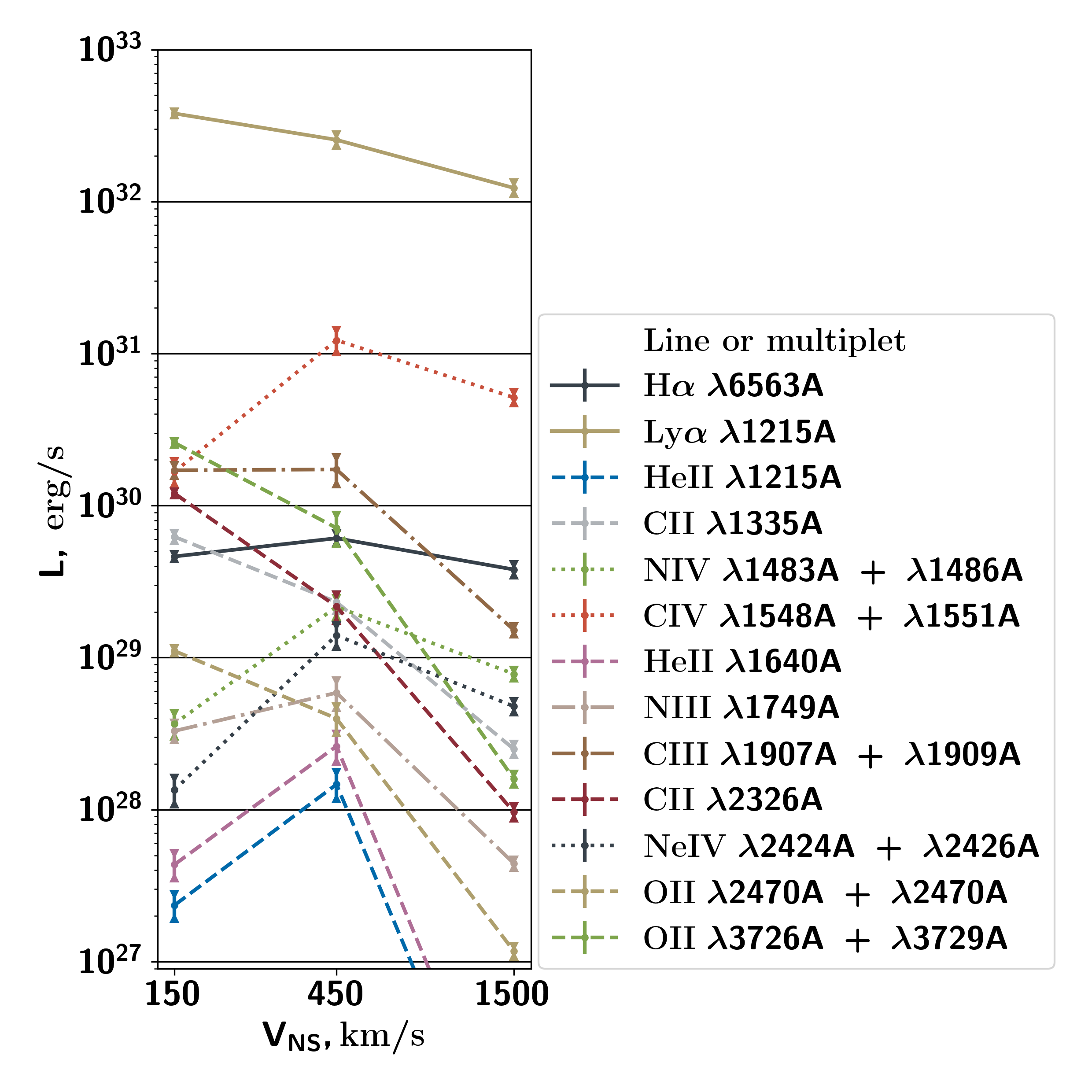}
        \caption{Model nebular luminosity, averaged over 100 years, as a function of pulsar velocity. Error bars indicate the standard deviation about the mean luminosity.}
        \label{fig:linesLuminosity}
\end{figure}

We analyse the ultraviolet lines using an approach previously applied successfully to optical transitions. The resulting luminosity--velocity relations are shown in Fig.~\ref{fig:linesLuminosity}. Their overall behaviour resembles that of the optical lines. The nebular line luminosity ($L$) is expected to scale with the pulsar wind power, which equals the neutron star's spin-down power ($\dot{E}$) \cite{2025PASA...42...79N, 2027JHEAp..5500710N}.

For singly ionized species (\CIIp{} and \OIIp{}), luminosity decreases monotonically with increasing velocity, with a knee in the relation. As the neutron star velocity increases from 150~km~s$^{-1}$ to 450~km~s$^{-1}$, the luminosity falls by a factor of 3 -- 5; at 1500~km~s$^{-1}$, it is more than an order of magnitude lower than at 150~km~s$^{-1}$. The decline is therefore expected to steepen towards high velocities, favouring observations of nebulae around relatively slow pulsars in these lines. The \HeIIpwne{} lines are an exception, exhibiting a peak in their luminosity--velocity relations. This difference is expected because the substantially higher ionization energy of helium affects the abundance of the emitting ions.

The lines of doubly ionized species (\CIIIp{} and \NIIIp{}) exhibit a plateau at low to moderate pulsar velocities, with comparable luminosities at $V_\text{NS} = 150$~km~s$^{-1}$ and 450~km~s$^{-1}$. At high velocities, the luminosity decreases by approximately a factor of ten. These lines therefore favour nebulae around pulsars with low to moderate velocities.

For triply ionized species (\CIVp, \NIVp{} and \NeIVp), the luminosity--velocity relations peak, as expected for optical lines. However, the ultraviolet luminosity rises by an order of magnitude between $V_\text{NS} = 150$~km~s$^{-1}$ and 450~km~s$^{-1}$, then falls by a factor of three at 1500~km~s$^{-1}$. Unlike most optical lines, the peak is expected to be steeper on the low-velocity side. These transitions thus favour observations of nebulae around relatively fast pulsars.

\subsubsection{Ultraviolet lines with the highest predicted luminosities}

The model ultraviolet lines are generally more luminous than their optical counterparts, by approximately an order of magnitude on average, making them promising observational targets. Below, we identify transitions with nebular luminosities exceeding $5 \times 10^{29}$~erg~s$^{-1}$ in the models that best reproduce the \HaLine{} observations.

Among singly ionized species, the brightest feature is \OIIlineLong{} (the 3726 and 3728~\AA{} doublet), with a luminosity of $3 \times 10^{30}$~erg~s$^{-1}$ at $V_\text{NS} = 150$~km~s$^{-1}$. In addition to its high luminosity, its location near the ultraviolet--optical boundary makes it particularly accessible: under suitable conditions, it can be observed with large ground-based facilities such as the 8-m VLT. Other bright transitions include \CIIlineShort{} and the 2326~\AA{} feature, with nebular luminosities of $1 \times 10^{30}$ and $6 \times 10^{29}$~erg~s$^{-1}$, respectively.

The brightest feature from a doubly ionized species is \CIIIline{} (the 1907 and 1909~\AA{} doublet). Its nebular luminosity reaches $2 \times 10^{30}$~erg~s$^{-1}$ over the neutron star velocity range $V_{\rm NS} = 150-450$~km~s$^{-1}$, making it a promising diagnostic of nebulae around pulsars with low to moderate velocities. As this range encompasses nearly all pulsars known to produce nebulae in the ISM, the feature could trace much of this nebular population in the ultraviolet.

The \CIVline{} doublet at 1548 and 1550~\AA{} is the brightest feature both among triply ionized species and among all the forbidden lines considered. Its luminosity reaches $1 \times 10^{31}$~erg~s$^{-1}$ at $V_\text{NS} = 450$~km~s$^{-1}$ and declines to $5 \times 10^{30}$~erg~s$^{-1}$ at 1500~km~s$^{-1}$. At $V_\text{NS} = 150$~km~s$^{-1}$, the predicted luminosity is $2 \times 10^{30}$~erg~s$^{-1}$. Among the ultraviolet forbidden lines considered, it is the only feature that permits relatively efficient observations of nebulae around high-velocity pulsars. Combined with its high luminosity, this makes it a particularly promising target for Spektr-UF.

Ly-$\alpha$ merits separate consideration. Model nebular luminosities reach $4 \times 10^{32}$~erg~s$^{-1}$ at $V_\text{NS} = 150$~km~s$^{-1}$, $3 \times 10^{32}$~erg~s$^{-1}$ at 450~km~s$^{-1}$ and $1 \times 10^{32}$~erg~s$^{-1}$ at 1500~km~s$^{-1}$, making it the brightest line considered. However, the nebula is optically thick in this transition, which may alter these predictions. Absorption by neutral hydrogen along the sightline may further affect its observability. All other transitions considered, apart from the hydrogen Lyman series, remain optically thin, including when Thomson scattering is taken into account.

Unlike plane-parallel shock models, detailed hydrodynamical models of nebulae around high-velocity pulsars account for spatial variations in flow velocity and the finite extent of the ionized ISM region. For the relatively slow pulsars J0437-4715, J2124-3358 and J1741-2054 \cite{2016ApJ...831..129R, 2017ApJ...835..264R, 2025A&A...696A.121A}, \CIVline{}, \CIIIline{}, \CIIlineShort{} and other transitions may contribute to the ultraviolet nebular emission detected by HST.

Unusual objects such as the Guitar Nebula around PSR J2225+6535 provide further motivation for these studies. A recent study \cite{2027JHEAp..5500710N} identified distinctive properties of this source and proposed a model for its formation. The model invokes non-equilibrium ionization at a bow shock propagating through very dense interstellar gas with a strong magnetic field, $B\sim 100$~$\mu$G, compressed by the radiative shock of an old supernova remnant. The nebula is anomalously bright given the relatively low spin-down power of J2225+6535. A reliable description of this source requires a multifluid treatment of the interaction between the ISM and the pulsar wind.

\section{The immediate environments of stars and protostars}

\subsection{Circumstellar dust} \label{sec:murga}

Interstellar dust plays a central role in many astrophysical processes. It regulates the thermal and ionization balance of the ISM and attenuates ultraviolet radiation within molecular clouds and hot cores, enabling complex organic molecules to form and survive. Grain surfaces also provide sites for molecular formation, from molecular hydrogen to these molecules. The smallest dust constituents---aromatic hydrocarbon macromolecules known as polycyclic aromatic hydrocarbons (PAHs)---are considered potential building blocks of prebiotic molecules.

Observations, laboratory experiments and theoretical calculations identify two broad components of interstellar dust: silicates and carbonaceous material. Carbonaceous dust can adopt diverse structures depending on its environment, directly affecting the physical and chemical properties of the grains.

Ultraviolet and infrared observations provide most of our information about dust properties. However, whereas the extensive infrared archive continues to grow through JWST observations, ultraviolet coverage remains comparatively limited. Spektr-UF could substantially expand this coverage and address three questions concerning carbonaceous dust and macromolecules:
\begin{itemize}
    \item[1.]{Characterizing dust structure and identifying carbonaceous nanoparticles during their early evolution around post-asymptotic giant branch (post-AGB) stars;}
    \item[2.]{Establishing links between the chemical composition of planetary nebulae (PNe) and their dust properties;}
    \item[3.]{Determining how the ultraviolet radiation field affects carbonaceous nanoparticle abundances.}
\end{itemize}

These objectives and the Spektr-UF instruments suitable for pursuing them are discussed below.

\subsubsection{Dust properties and carbonaceous nanoparticles around post-AGB stars}

The infrared spectra of carbon-rich post-AGB stars exhibit diverse features. In addition to continuum emission beyond 30~$\mu$m, they contain numerous bands between 3 and 30~$\mu$m. Some, including those at 7.7 and 11.3~$\mu$m, have been associated with specific carriers, such as PAHs or silicon carbide. Candidate carriers have been proposed for other bands, but definitive identifications remain elusive.

Most carbonaceous dust forms in the envelopes of AGB stars \cite{2009ApJ...698.1136G}. Yet the characteristic infrared PAH bands are absent from most single AGB stars, probably for two reasons. First, grains in their dense envelopes appear to grow rapidly to micrometre sizes or larger, losing the spectral signatures characteristic of nanoparticles. Second, efficient PAH excitation generally requires an external ultraviolet source, which is absent around an isolated AGB star. Although PAHs can absorb optical photons~\cite{li02}, this excitation is probably insufficient to produce bright PAH bands within a dense AGB envelope.

Emission from PAHs and other nanoparticles becomes apparent at later evolutionary stages: during the post-AGB phase, when the hot central source---the emerging white dwarf---supplies ultraviolet radiation, and in binary systems where a hot companion illuminates the AGB envelope~\cite{smolders10}. Infrared spectra are particularly informative at these early stages of dust evolution, when the material is relatively unprocessed and retains clues to its formation.

Ultraviolet observations of post-AGB stars already identified as infrared sources would provide complementary constraints on nanoparticle abundances and morphology. Two approaches are possible: characterizing the slope of the extinction curve and the 2175~\AA{} bump, and searching for weak absorption bands associated with electronic transitions in specific carbonaceous structures.

Many carbonaceous particles exhibit characteristic absorption near 2175~\AA{}. The position and width of this feature can vary with particle geometry, size and morphology~\cite{steglich10, papoular13}. Comparing infrared PAH emission bands with the position, width and strength of the ultraviolet bump in the same objects, such as post-AGB stars observed in both wavelength ranges, would be particularly valuable. Such comparisons could connect dust absorption and emission properties and constrain the nature of newly formed carbonaceous nanoparticles.

\begin{figure}
    \centering
    \includegraphics[width=0.95\linewidth]{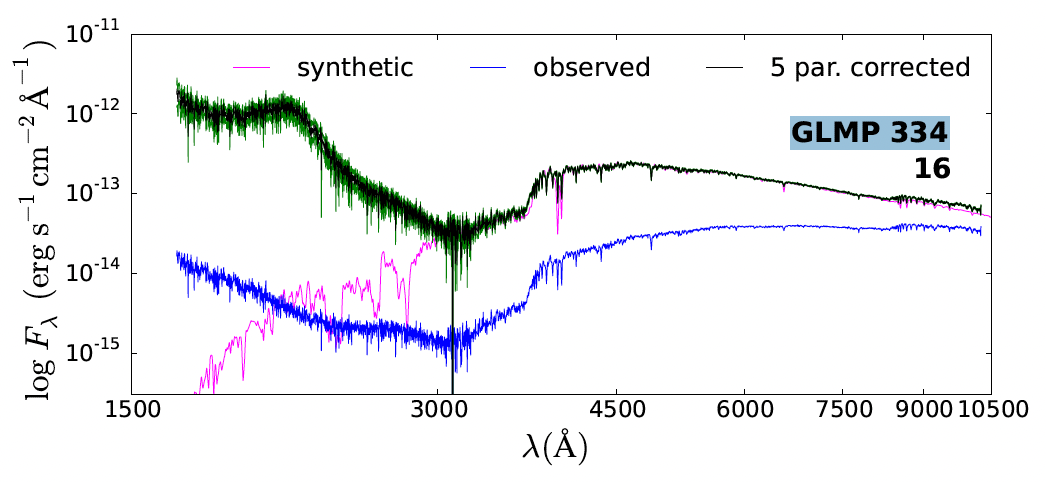}
    \caption{HST/STIS spectrum of the post-AGB star GLMP 334. The observed spectrum is shown in blue, the synthetic spectrum in magenta and the extinction-corrected spectrum in black. Green indicates the uncertainty envelope of the corrected spectrum. Reproduced from \cite{2018A&A...615A.115K} in accordance with the publisher's permissions.}
    \label{fig:murga}
\end{figure}

In addition to the strong 2175~\AA{} feature, ultraviolet absorption spectra contain weaker, broad bands characteristic of different particle types. Sufficient spectral resolution and sensitivity could allow these features to be detected and identified using laboratory spectral databases, such as \href{http://astrochemistry.ca.astro.it/database/pahs.html}{http://astrochemistry.ca.astro.it/database/pahs.html}. Figure~\ref{fig:murga} shows an HST/STIS spectrum of the post-AGB star GLMP 334~\cite{2018A&A...615A.115K}. Its relatively low temperature, 7400~K, places it at an early post-AGB stage. The ultraviolet radiation field is therefore weak, and the dust should retain a structure close to that established during its formation, having undergone little ultraviolet processing. High-resolution, high-sensitivity spectra of such stars could illuminate the earliest stages of interstellar dust evolution.

Together, these approaches would constrain both the ensemble properties of dust, encoded in the extinction curve, and the specific molecular and nanoscale carbonaceous structures present in the ISM.

An extensive infrared dataset already exists for post-AGB stars in the Milky Way and the Magellanic Clouds~\cite{peeters02, sloan14}, and continues to expand through JWST observations. Objects with existing infrared spectra are natural candidates for Spektr-UF observations. Because post-AGB stars are faint in the ultraviolet, the feasibility of obtaining useful spectra must be assessed carefully. Even a small number of spectra of stars at the earliest post-AGB stages, in either the Milky Way or the Magellanic Clouds, would constrain the spectral signatures of newly formed dust.

\subsubsection{Links between planetary nebula chemistry and dust properties}

Developing a theory of dust formation in stellar envelopes and connecting dust properties to envelope chemistry requires elemental abundance measurements. For carbonaceous dust, the carbon-to-hydrogen and carbon-to-oxygen ratios (C/H and C/O) are particularly important. Their possible relationship to the dust composition of PNe---aromatic, aliphatic, mixed or fullerene-bearing---is of special interest.

Most diagnostic carbon lines lie in the ultraviolet, including the prominent CIV$\lambda$~1550~\AA{}, CIII]$\lambda$~1909~\AA{} and CII]$\lambda$~2325-29~\AA{} features. Ultraviolet observations are difficult during the AGB and post-AGB phases, but envelope abundances can be measured at the subsequent PN stage, when these objects become bright ultraviolet sources. Because carbon diagnostics lie primarily in the ultraviolet, its abundance is less well characterized than those of other elements. Published HST ultraviolet spectra cover several dozen PNe, but only a few of these also have infrared observations~\cite{ventura17, stanghellini22}. Establishing correlations between nebular chemistry and dust properties requires a substantially larger sample observed in both wavelength ranges.

Unlike the preceding objective, this study requires the high-resolution UVES spectrograph. Suitable targets include PNe from the samples of \cite{garcia14} and \cite{ventura22} that either lack ultraviolet spectra or have only poor-quality data; eighteen such objects are known. One example is Tc~1~\cite{aleman19}, where fullerenes were first identified~\cite{cami10} and are particularly abundant. Existing IUE spectra of this PN~\cite{pottasch11} are of limited quality, making it a promising target for Spektr-UF.

\subsubsection{The effects of ultraviolet radiation on carbonaceous nanoparticles}

Carbonaceous nanoparticles are sensitive to ultraviolet radiation. PAHs form one subset of this population; hereafter, we use PAHs as shorthand that also encompasses other carbonaceous nanoparticles containing aromatic structures. Absorption of photons with energies as low as 3--4~eV heats these particles and excites their vibrational modes. Intense ultraviolet fields in star-forming regions can heat PAHs to high temperatures, making these regions prominent sites of PAH emission in the Milky Way and other galaxies. PAH emission is therefore used as a star formation rate indicator in both the nearby and distant Universe~\cite{calzetti07, gregg24, pathak24, shivaei24}.

Absorption of energetic photons can also ionize and dissociate PAHs, substantially altering their abundances, sizes, structures and charge states. These interactions have been investigated through several complementary approaches. Observational links between radiation-field properties and PAH characteristics have been studied in individual objects, such as PDRs~\cite{murga22}, and in large samples of extragalactic star-forming regions~\cite{khramtsova14, egorov23, egorov25}. Laboratory experiments reveal complex interactions with multiple possible pathways~\cite{ekern98, zhen14, zhen_2016, wenzel20}, while theoretical studies model the underlying processes~\cite{allain96, lepage01, murga16, andrews16}. Despite this extensive work, a complete, quantitative description of interstellar PAH processing by ultraviolet radiation remains elusive.

A central limitation is the lack of direct comparisons between ultraviolet radiation-field intensity and hardness on the one hand, and PAH properties on the other. PAH abundances are systematically lower in star-forming regions than in the diffuse ISM~\cite{chastenet19, aniano20, egorov25}, indicating destruction. A study by \cite{khramtsova13} found no correlation between PAH abundance and radiation-field intensity in HII regions. However, the intensity was inferred by modelling the dust emission, including that of the PAHs themselves, making it an indirect estimate subject to additional uncertainties. Studies by \cite{khramtsova14, wiebe14} compared PAH abundances with the radiation-hardness indicator $\rm [OIII]/H_{\beta}$, but this ratio also depends on other properties, including metallicity. More recently, \cite{egorov23, egorov25} found that PAH abundance correlates with the sulphur line ratio $\rm [SIII]/[SII]$, interpreted in this context as a tracer of the ionization parameter and hence the flux of hydrogen-ionizing photons with energies above 13.6~eV. They concluded that extreme-ultraviolet photons (E$>$ 13.6~eV) dominate PAH destruction, whereas radiation hardness---the spectral distribution within the ultraviolet---plays a minor role. The absence of correlations between PAH abundance and radiation-hardness indicators supports this interpretation.

Individual objects, however, present a more complex picture. In the Orion HII region and the adjoining Orion Bar PDR, variations in radiation-field intensity are accompanied by changes in PAH size, charge state and hydrogen-to-carbon ratio. The conclusion that non-hydrogen-ionizing photons with energies of 3--13.6~eV play no role in PAH evolution therefore requires closer examination with additional observations. Moreover, the extragalactic HII regions studied by \cite{egorov23, egorov25} are generally unresolved into individual stars. Each observed region includes both ionized gas around stellar clusters and surrounding PDRs, and often encompasses multiple clusters, PDRs and diffuse gas. Because PAH emission originates in PDRs \cite{2013ARep...57..573P, 2024A&A...685A..75C}, its properties must reflect the local radiation field, including its non-ionizing ultraviolet component.

A study of three galaxies by \cite{ujjwal24} compared PAH-band fluxes from star-forming regions with far-ultraviolet measurements obtained through the 1480\AA{} filter of the Ultraviolet Imaging Telescope aboard AstroSat (AstroSat/UVIT). Regions with high star formation rates contained larger fractions of ionized and small PAHs. This finding suggests that energetic ultraviolet radiation, potentially including photons in the 3--13.6~eV range, modifies PAH properties.

A more comprehensive and statistically robust assessment requires comparisons of PAH properties with ultraviolet intensities at several wavelengths, together with ultraviolet colour ratios, across a large sample of star-forming regions spanning different metallicities, sizes and other physical properties. Spektr-UF could provide these measurements through near- and far-ultraviolet photometry using broad-band filters in its field camera unit, with different central wavelengths. Its wide field of view, high angular resolution and sensitivity would enable the first multiband survey of nearby galaxies spanning the full ultraviolet range at this combination of capabilities, beyond those of earlier instruments such as the Solar Blind Channel of Hubble's Advanced Camera for Surveys (HST/ACS/SBC) and AstroSat/UVIT. Suitable targets include PHANGS galaxies with existing complementary JWST and VLT observations presented in \cite{egorov25}.

\subsection{Protoplanetary discs}\label{sec:disks}

A defining property of classical T Tauri stars, which ultimately helped establish their nature as young stars surrounded by protoplanetary discs, is excess blue and ultraviolet emission \cite{1955BOTT....2l..33H}, powered primarily by accretion onto the star. This excess has long made T Tauri stars attractive targets for ultraviolet observations \cite{1998ESASP.413...59G}. Such studies have three principal motivations: 1)~ultraviolet spectral lines diagnose conditions in accretion flows and warm disc atmospheres; 2)~ultraviolet radiation heats the disc, strongly influencing its physical structure and evolution, including photoevaporation; and 3)~ultraviolet radiation drives photochemistry in the disc surface layers, shaping their chemical composition. Ultraviolet studies therefore require both high-resolution observations of individual lines and measurements of the overall ultraviolet spectral energy distribution. The potential of Spektr-UF observations of hot-gas tracers, including CIV 1550\AA, SiIV 1400\AA, NV 1240\AA, OVI 1035\AA{} and other lines in T Tauri spectra, was previously discussed in \cite{2011Ap&SS.335...61G}. Here, we examine several complementary objectives.

Lines of \hh{} and CO are securely detected in protoplanetary discs, appearing in emission or absorption depending on the viewing geometry. Fluorescent \hh{} emission pumped by Ly$\alpha$ is observed in many classical T Tauri stars and originates partly in disc winds and partly in the disc itself \cite{2006ApJS..165..256H,2023AJ....166...67F,2025AJ....169..240K}. Disc emission allows ultraviolet \hh{} lines to trace radial variations in gas properties across the terrestrial planet-forming region and out to $\sim10$~au \cite{2023AJ....166...67F}. Combining these measurements with CO observations probing the same spatial scales constrains the CO/\hh{} ratio across the disc. This ratio is fundamental because CO abundance inferred from radio observations is used to estimate the total disc gas mass \cite{2017ApJ...849..130M}, commonly assuming the mean interstellar CO/\hh{} ratio of $10^{-4}$. Fluorescence observations, however, indicate that this ratio substantially exceeds the interstellar value in some objects, approaching values measured in cometary material. In the discs around V~4046~Sgr, HN~Tau and RECX-11, for example, CO/\hh{} is $\approx 1$, potentially indicating photoevaporation and/or collisional destruction of large grains or planetesimals \cite{2011ApJ...734...31F}. In AA~Tau, where \hh{} and CO were observed in absorption, CO/\hh~$\approx 0.4$ \cite{2012ApJ...744...22F}. These high values may partly reflect differences between the regions traced by CO and \hh{}. Observations of a statistically meaningful sample would refine the inferred CO/\hh{} ratios and establish their dependence on the spectral type of the central star and on disc mass, age and evolutionary state.

Fluorescent molecular hydrogen lines can also constrain extinction towards protoplanetary discs, an essential input to the interpretation of many observations. Radiative-transfer modelling of these lines can recover the shape of the interstellar extinction curve over approximately 1150--1700\AA{} \cite{2016ApJ...828...69M}. Assuming that the lines are optically thin simplifies the analysis by avoiding detailed radiative-transfer calculations. Using this approximation, authors of \cite{2024A&A...692A..69F} determined $A_V$ and $R_V$ towards 34 T Tauri stars by comparing observed and predicted intensities of strong ultraviolet \hh{} lines, finding good agreement with extinction estimates from optical data.

Another feature of T Tauri disc spectra, superimposed on the fluorescent emission, is the 1600~\AA{} bump, detected in several dozen discs \cite{2017ApJ...844..169F, 2023AJ....166...67F}. Initially attributed to continuum emission from electron-impact-excited \hh{} \cite{2004ApJ...607..369H,2004ApJ...614L.133B}, this mechanism was subsequently shown to be unable to reproduce the observed properties of the feature \cite{2017ApJ...844..169F}. A better match is obtained if the emission arises from \hh{} produced by Ly$\alpha$ photodissociation of water. If correct, this interpretation implies that water survives at distances of 1--2~au from the star despite intense ultraviolet irradiation. In the GM~Aug disc, for example, the 1600~\AA{} bump can be explained by Ly$\alpha$-driven water dissociation at a rate of $1.7\cdot10^{42}$~s$^{-1}$. JWST mid-infrared observations confirm the presence of water and hydroxyl in this disc \cite{2025ApJ...991..128R}, although they imply a lower water dissociation rate of $5.4\cdot10^{40}$~s$^{-1}$. Spektr-UF studies of discs exhibiting the 1600~\AA{} bump should build on the JWST constraints on disc water abundances available by the time of launch.

Accurate calculations of disc physical and chemical structure require knowledge of the spectral energy distribution illuminating the disc. Thermochemical models (for example, \cite{2009A&A...501..383W,2013ApJ...766....8A}) determine disc properties through coupled calculations of thermal balance and radiative transfer. The central radiation field may be represented by a simple sum of stellar and accretion-region blackbody spectra, or by a more realistic, scaled spectrum of an observed T Tauri star. The adopted ultraviolet spectral shape can produce substantial differences in the modelled temperature and density distributions \cite{2011Ap&SS.335...33A}, with corresponding consequences for predictions of disc chemistry and molecular line profiles.

Modelling individual systems therefore benefits from measurements of their ultraviolet spectra. Although many ultraviolet spectra of T Tauri stars have been published \cite{2000ApJS..129..399V,2004ApJ...614L.133B,2012ApJ...744..121Y,2014ApJ...784..127F}, coverage remains particularly sparse for young brown dwarfs with protoplanetary discs. Only a few ultraviolet spectra of accreting brown dwarfs are available \cite{2005ApJ...630L..89G,2010ApJ...715..596F,2026ApJ...999..111P}, and two brown dwarfs are included in the sample of \cite{2012ApJ...744..121Y}. These spectra suggest that, at low masses and accretion rates, disc properties may depart from a simple extrapolation of those of more massive systems. Accreting brown dwarfs therefore represent promising targets for Spektr-UF.

\section*{Funding}

The sections ``The interstellar medium of the Milky Way and molecular clouds'' and ``Protoplanetary discs'' were prepared with contributions from M.~S.~Kirsanova and A.~D.~Yarovova, supported by BASIS Foundation grants 25-1-2-88-1 and 25-1-2-88-4. The section ``Nebulae around high-velocity pulsars'' was prepared by M.~V.~Barkov and I.~N.~Nikonorov, supported by BASIS Foundation grant 24-1-2-25-1. The section ``Small-scale structure in interstellar clouds'' was prepared by I.~I.~Zinchenko, supported by Russian Science Foundation grant 24-12-00153.

\def\aj{{AJ}}                   
\def\actaa{{Acta Astron.}}      
\def\araa{{ARA\&A}}             
\def\apj{{ApJ}}                 
\def\apjl{{ApJL}}                
\def\apjs{{ApJS}}               
\def\ao{{Appl.~Opt.}}           
\def\apss{{Ap\&SS}}             
\def\aap{{A\&A}}                
\def\aapr{{A\&A~Rev.}}          
\def\aaps{{A\&AS}}              
\def\azh{{AZh}}                 
\def\baas{{BAAS}}               
\def\bac{{Bull. astr. Inst. Czechosl.}}
\def\caa{{Chinese Astron. Astrophys.}}
\def\cjaa{{Chinese J. Astron. Astrophys.}}
\def\icarus{{Icarus}}           
\def\jcap{{J. Cosmology Astropart. Phys.}}
\def\jrasc{{JRASC}}             
\def\memras{{MmRAS}}            
\def\mnras{{MNRAS}}             
\def\na{{New A}}                
\def\nar{{New A Rev.}}          
\def\pra{{Phys.~Rev.~A}}        
\def\prb{{Phys.~Rev.~B}}        
\def\prc{{Phys.~Rev.~C}}        
\def\prd{{Phys.~Rev.~D}}        
\def\pre{{Phys.~Rev.~E}}        
\def\prl{{Phys.~Rev.~Lett.}}    
\def\pasa{{PASA}}               
\def\pasp{{PASP}}               
\def\pasj{{PASJ}}               
\def\rmxaa{{Rev. Mexicana Astron. Astrofis.}}
\def\qjras{{QJRAS}}             
\def\skytel{{S\&T}}             
\def\solphys{{Sol.~Phys.}}      
\def\sovast{{Soviet~Ast.}}      
\def\ssr{{Space~Sci.~Rev.}}     
\def\zap{{ZAp}}                 
\def\nat{{Nature}}              
\def\iaucirc{{IAU~Circ.}}       
\def\aplett{{Astrophys.~Lett.}} 
\def\apspr{{Astrophys.~Space~Phys.~Res.}}
\def\bain{{Bull.~Astron.~Inst.~Netherlands}} 
\def\fcp{{Fund.~Cosmic~Phys.}}  
\def\gca{{Geochim.~Cosmochim.~Acta}}   
\def\grl{{Geophys.~Res.~Lett.}} 
\def\jcp{{J.~Chem.~Phys.}}      
\def\jgr{{J.~Geophys.~Res.}}    
\def\jqsrt{{J.~Quant.~Spec.~Radiat.~Transf.}}
\def\memsai{{Mem.~Soc.~Astron.~Italiana}}
\def\nphysa{{Nucl.~Phys.~A}}   
\def\physrep{{Phys.~Rep.}}   
\def\physscr{{Phys.~Scr}}   
\def\planss{{Planet.~Space~Sci.}}   
\def\procspie{{Proc.~SPIE}}   

\let\astap=\aap
\let\apjlett=\apjl
\let\apjsupp=\apjs
\let\applopt=\ao

\bibliographystyle{ufn}
\bibliography{uv-ism,uv-ism-wiebe,uv-ism-barkov-nikonorov,uv-ism-balashev,library}

\newpage

\end{document}